\documentclass[12pt, oneside]{amsproc}

\usepackage{amsthm,amsmath,amssymb,amsfonts,amscd,mathtools,upgreek,url,
}
\usepackage[foot]{amsaddr}

\newcommand{\cS}{{\mathbf{S}}}
\newcommand{\tdt}{\mathbf{D}_t}

\newcommand{\bc}{y}
\newcommand{\pe}{u}
\newcommand{\vv}{v}

\newcommand{\mv}{w}
\newcommand{\iup}{\upmu}

\newcommand{\fint}{\mathrm{b}}
\newcommand{\secint}{\mathrm{c}}

\newcommand{\hc}{\upsigma}
\newcommand{\td}{\mathrm{s}}

\newcommand{\ik}{k}

\newcommand{\cto}{\uprho_1}
\newcommand{\ctt}{\uprho_2}

\newcommand{\mme}{\mathbf{m}}
\newcommand{\lte}{\mathbf{w}}

\newcommand{\ap}{\upalpha}

\newcommand{\tdnew}{\upxi}

\newcommand{\ud}{u_\ik,\dots}
\newcommand{\ua}{{\mathrm{p}}}
\newcommand{\ub}{{\mathrm{q}}}
\newcommand{\tua}{{\tilde{\ua}}}
\newcommand{\tub}{{\tilde{\ub}}}

\newcommand{\fn}{\upvarphi}
\newcommand{\gnew}{\uppsi}

\newcommand{\mf}{\Omega}
\newcommand{\la}{\uplambda}

\newcommand{\ff}{\mathbf{Q}}
\newcommand{\tff}{\tilde{\ff}}

\newcommand{\fik}{\mathbb{C}}

\newcommand{\sm}{\mathrm{d}}

\newcommand{\lb}{\label}
\newcommand{\er}{\eqref}

\newcommand{\zp}{\mathbb{Z}_{\ge 0}}
\newcommand{\zsp}{\mathbb{Z}_{>0}}
\newcommand{\zz}{\mathbb{Z}}

\newcommand{\pd}{\partial}
\newcommand{\lt}{\mathbf{W}}

\newcommand{\mm}{\mathbf{M}}

\newcommand{\diunew}{\mathrm{L}}

\newtheorem{proposition}{Proposition}

\theoremstyle{definition}
\newtheorem{definition}{Definition}
\newtheorem{example}{Example}
\newtheorem{remark}{Remark}

\begin{document}


\title[Lax representations and 
constructions of Miura-type transformations]{On matrix Lax representations and 
constructions of Miura-type transformations for differential-difference equations}
\date{}

\author{Evgeny Chistov$^1$ and Sergei Igonin$^1$} 
\address{$^1$Center of Integrable Systems, P.G.~Demidov Yaroslavl State University, Yaroslavl, Russia} 
\email{e-chistov01@yandex.ru, s-igonin@yandex.ru}





\begin{abstract}

This paper is part of a research project on relations between 
differential-difference matrix Lax representations (MLRs) with the action of gauge transformations 
and discrete Miura-type transformations (MTs) 
for (nonlinear) integrable differential-difference equations.

The paper addresses the following problem:

\noindent 
when and how can one eliminate some shifts 
of dependent variables from the discrete (difference) part of an MLR by means of 
gauge transformations?

Using results on this problem, we present applications to constructing new MLRs 
and new integrable equations connected by new MTs to known equations.

In particular, we obtain results of this kind for equations connected to the following ones:
\begin{itemize}
	\item The two-component Belov--Chaltikian lattice.
	\item The equation (introduced by G.~Mar\'i Beffa and Jing~Ping~Wang)
	which describes the evolution induced on invariants by an invariant evolution of planar polygons.
	\item An equation from~[S.~Igonin,~arXiv:2405.08579] related to the
	Toda lattice in the Flaschka--Manakov coordinates. 
\end{itemize}
\end{abstract}

\maketitle

\section{Introduction}
\lb{secint}

This paper is part of a research project on relations between 
differential-difference matrix Lax representations (MLRs) with the action of gauge transformations 
and discrete Miura-type transformations, 
which belong to the main tools in the theory of (nonlinear) integrable differential-difference equations.
Such equations occupy a prominent place in the theory of integrable systems.
In particular, equations of this type arise as discretizations of integrable 
partial differential equations (PDEs) and various geometric constructions
and as chains associated with Darboux and B\"acklund transformations of PDEs 
(see, e.g.,~\cite{BeffaWang2013,HJN-book16,kmw,LWY-book22} and references therein).

In this paper, $\zsp$ and $\zp$ denote the sets of positive and nonnegative integers, respectively.
Let $\diunew\in\zsp$. 
As explained in Section~\ref{secprel}, we consider an evolutionary differential-difference equation 
for an $\diunew$-component vector-function $u=\big(u^1(n,t),\dots,u^{\diunew}(n,t)\big)$, 
where $n$ is an integer variable and $t$ is a real or complex variable.
Thus the components $u^1,\dots,u^{\diunew}$ of~$u$ are the dependent variables in the equation.

For each fixed integer~$\ik$, we consider also the vector-function
$u_\ik=u_\ik(n,t)$ given by the formula $u_\ik(n,t)=u_\ik(n+\ik,t)$.
The vector-functions~$u_\ik=(u^1_\ik,\dots,u^{\diunew}_\ik)$, $\ik\in\zz$, 
are called the \emph{shifts} of~$u=(u^1,\dots,u^{\diunew})$.

In Definition~\ref{dmlpgtnew} in Section~\ref{secprel}
we recall the notions of matrix Lax representations (MLRs) for 
evolutionary differential-difference equations and 
an action of the group of matrix gauge transformations
on the set of all MLRs of a given equation.
As discussed in Definition~\ref{dmlpgtnew}, in an MLR $(\mm,\,\lt)$ 
the matrix~$\mm$ can be called the discrete (difference) part of the MLR.
For brevity, we call~$\mm$ the \emph{$\cS$-part} of the MLR.
Here $\cS$ is the shift operator defined in Section~\ref{secprel}.

In general, $\mm$ depends on a finite number of the dynamical
variables~$u_{\ik}=(u^1_{\ik},\dots,u^\diunew_{\ik})$, $\,\ik\in\zz$, 
and a parameter~$\la$. For any fixed integers $r_1,\dots,r_\diunew$, we can relabel 
\begin{gather}
\lb{relabnew}
u^1_\ik\,\mapsto\,u^1_{\ik+r_1},\,\quad\dots\,\quad,\,\quad 
u^{\diunew}_\ik\,\mapsto\,u^{\diunew}_{\ik+r_\diunew}\qquad\quad\forall\,\ik\in\zz.
\end{gather}
Relabeling~\er{relabnew} says that in the considered differential-difference equation
one makes the invertible change of variables 
\begin{gather}
\notag
u^1(n,t)\,\mapsto\,u^1(n+r_1,t),\,\quad\dots\,\quad,\,\quad 
u^{\diunew}(n,t)\,\mapsto\,u^{\diunew}(n+r_\diunew,t).
\end{gather}
Applying a relabeling~\er{relabnew} with suitable fixed $r_1,\dots,r_\diunew\in\zz$, 
we can transform~$\mm$ to the form
$\mm=\mm(u_0,\dots,u_{\mathrm{a}},\la)$ for some $\mathrm{a}\in\zp$.

This paper deals mainly with MLRs $(\mm,\,\lt)$ such that the $\cS$-part~$\mm$
may depend only on~$u_0,\,u_1,\,u_2,\,\la$. Thus $\mm$ is of the form
\begin{gather}
\lb{muu2}
\mm=\mm(u_0,u_1,u_2,\la).
\end{gather}
In the existing literature on differential-difference equations
with two independent variables $n$ and $t$, 
the majority of known MLRs are of the type~\er{muu2} 
or can be brought to the form~\er{muu2} 
by a relabeling~\er{relabnew} with suitable $r_1,\dots,r_\diunew$.

In~\er{muu2} one does not require for~$\mm$ nontrivial dependence on all~$u_0,\,u_1,\,u_2$.
For instance, in some of the examples considered in this paper the matrix~$\mm$ 
does not depend on~$u_2$.

We use a result from~\cite{IgLaxGauge2024} 
(formulated in Proposition~\ref{tmlr} in Section~\ref{secprel} of the present paper) 
describing sufficient conditions which allow one 
to eliminate the dependence on~$u_1,\,u_2$ in the~\mbox{$\cS$-part} 
of a given MLR by means of a gauge transformation.
As discussed below, 
in the present paper we demonstrate how this result helps to construct
new MLRs and new integrable equations connected by new MTs to known equations.
We give detailed descriptions of such constructions in concrete examples.
The constructions presented in this paper can be applied to many more examples,
which will be considered elsewhere.

When one makes a classification for some class of (1+1)-dimensional integrable equations 
(which may be partial differential, difference or differential-difference), 
it is often possible to obtain an explicit list of basic equations such that all the other equations 
from the considered class can be derived from the basic ones by using~MTs
(see, e.g.,~\cite{mss91,yam2006,garif2018,meshk2008,LWY-book22} and references therein).
Furthermore, MTs often help to construct conservation laws as well as 
Darboux and B\"acklund transformations for (1+1)-dimensional equations 
(see, e.g.,~\cite{adler-pos11,miura-cl,suris03,svinin11}).
Hence it is desirable to develop constructions of MTs.

In this paper we present examples of constructions of MTs in the situation 
when one has an equation~$E$ possessing an MLR $(\mm,\,\lt)$ 
with $\cS$-part $\mm=\mm(u_0,u_1,u_2,\la)$ satisfying conditions of Proposition~\ref{tmlr}.
In such a situation Proposition~\ref{tmlr} allows us 
to transform (by means of a gauge transformation) the MLR $(\mm,\,\lt)$ to a new MLR 
$(\hat{\mm},\,\hat{\lt})$ with $\cS$-part of the form $\hat{\mm}=\hat{\mm}(u_0,\la)$.
Then we construct MTs by exploring the structure of the auxiliary linear system 
associated with the MLR~$(\hat{\mm},\,\hat{\lt})$.

Using this approach, in Section~\ref{sbc} we construct the following.
\begin{itemize}
	\item A parametric family of new integrable equations~\er{dtv} 
	depending on a parameter~$\ap\in\fik$.
	\item A parametric family of new MTs~\er{cmts} from~\er{dtv} to 
	the two-component Belov--Chaltikian lattice~\er{bce}, 
	which is the Boussinesq lattice related to the lattice $W_3$-algebra~\cite{Belov93,hiin97,BeffaWang2013,kmw}.
	\item A parametric family of new MTs~\er{mtpeqq} from~\er{dtv} to equation~\er{peq} 
	introduced in~\cite{BeffaWang2013}. As shown by G.~Mar\'i Beffa and Jing~Ping~Wang~\cite{BeffaWang2013}, the integrable equation~\er{peq} describes the evolution induced on invariants 
by an invariant evolution of planar polygons.
\end{itemize}

Also, we obtain a new MLR~\er{rchmu} for equation~\er{peq} 
and a new MLR (described in Remark~\ref{rintdtv}) for equation~\er{dtv}.
Each of these MLRs has a parameter~$\la\in\fik$.

In Section~\ref{stoda} we consider the integrable equation~\er{mtoda}
constructed in~\cite{IgLaxGauge2024} with the MT~\er{dstoda} 
to the Toda lattice~\er{todanew} written in the Flaschka--Manakov coordinates.
As shown in~\cite{IgLaxGauge2024}, equation~\er{mtoda} possesses 
the MLR~\er{Mmtoda},~\er{Umtoda} with 
$\cS$-part of the form $\mm=\mm(u_{0},u_{1},\la)$.
Proposition~\ref{tmlr} is applicable here, 
but in this case the straightforward application of Proposition~\ref{tmlr} 
gives rather cumbersome formulas.

Because of this, in Section~\ref{stoda} we present a different approach, 
in order to eliminate the dependence on~$u_1$ in the~\mbox{$\cS$-part} 
$\mm(u_{0},u_{1},\la)$ by means of a gauge transformation.
We find a suitable gauge transformation by solving some matrix ordinary differential equations.

Using the obtained gauge equivalent MLR with \mbox{$\cS$-part} 
depending only on~$u_{0}$ and $\la$, in Section~\ref{stoda} we construct 
\begin{itemize}
	\item a parametric family of new integrable equations~\er{eqhw} 
	depending on a parameter~$\hc\in\fik$;
	\item new MTs~\er{mthw},~\er{tdw} from~\er{eqhw} 
	to known equations~\er{mtoda},~\er{todanew} related to the Toda lattice;
	\item a new MLR \er{tMmtoda}, \er{tUmtoda} for equation~\er{mtoda};
	\item a new MLR (described in Remark~\ref{rinteqhw}) for equation~\er{dtv}.
\end{itemize}
Each of the constructed MLRs has a parameter~$\la\in\fik$.

\section{Preliminaries}
\lb{secprel}

In this paper scalar functions are assumed to take values in~$\fik$.

Fix $\diunew\in\zsp$. Below we consider an equation 
for an $\diunew$-component vector-function 
$$
u=\big(u^1(n,t),\dots,u^{\diunew}(n,t)\big)
$$
of an integer variable~$n$ and a real or complex variable~$t$.
For any fixed integer~$\ik$, 
we have $u_\ik=(u^1_\ik,\dots,u^{\diunew}_\ik)$,
where for each $\iup=1,\dots,\diunew$ 
the component $u_\ik^\iup$ is a function
of $n,t$ given by the formula $u_\ik^\iup(n,t)=u^\iup(n+\ik,t)$.
In particular, $u_0=u$.

Let $\ua,\ub\in\zz$ such that $\ua\le\ub$.
Consider an evolutionary differential-difference equation
\begin{gather}
\lb{sddenew}
\pd_t(u)=\ff(u_\ua,u_{\ua+1},\dots,u_\ub),
\end{gather}
where $\ff$ is an $\diunew$-component vector-function $\ff=(\ff^1,\dots,\ff^{\diunew})$.

The differential-difference equation \eqref{sddenew} says that one has 
the following infinite set of differential equations
$$
\pd_t\big(u(n,t)\big)=\ff\big(u(n+\ua,t),u(n+\ua+1,t),\dots,u(n+\ub,t)\big),\qquad\quad n\in\zz.
$$
Considering the components of $u=\big(u^1(n,t),\dots,u^{\diunew}(n,t)\big)$ 
and $\ff=(\ff^1,\dots,\ff^{\diunew})$, we can rewrite~\eqref{sddenew} as
\begin{gather}
\lb{msdde}
\pd_t\big(u^i\big)=\ff^i(u_\ua^\iup,u_{\ua+1}^\iup,\dots,u_\ub^\iup),\qquad\quad
i=1,\dots,\diunew,
\end{gather}
which yields 
\begin{gather}
\lb{uildde}
\pd_t\big(u^i_\ik\big)=\ff^i(u_{\ua+\ik}^\iup,u_{\ua+1+\ik}^\iup,\dots,u_{\ub+\ik}^\iup),\qquad\quad
i=1,\dots,\diunew,\qquad \ik\in\zz.
\end{gather}

We employ the formal theory of evolutionary differential-difference equations, which views 
\begin{gather}
\lb{uldiunew}
u_\ik=(u^1_\ik,\dots,u^\diunew_\ik),\qquad\quad \ik\in\zz,
\end{gather}
as independent quantities called \emph{dynamical variables}.
In what follows, the notation $\fn=\fn(\ud)$ says 
that a function~$\fn$ depends on a finite 
number of the dynamical variables $u_\ik^{\iup}$ for $\ik\in\zz$ and $\iup=1,\dots,\diunew$.

For any fixed integers $\fint\le\secint$, 
the notations $\fn=\fn(u_\fint,\dots,u_\secint)$ and $\fn=\fn(u_\fint,u_{\fint+1},\dots,u_\secint)$
indicate that $\fn$ may depend on~$u_\ik^{\iup}$ with $\ik=\fint,\dots,\secint$ 
and $\iup=1,\dots,\diunew$.

Let $\cS$ be the \emph{shift operator} with respect to the variable~$n$.
Applying~$\cS$ to a function $\gnew=\gnew(n,t)$, one gets the function~$\cS(\gnew)$
such that $\cS(\gnew)(n,t)=\gnew(n+1,t)$. Clearly, the operator~$\cS$ is invertible,
which allows us to consider, for each integer~$\ell$, 
the $\ell$th power~$\cS^\ell$ of~$\cS$. One has $\cS^\ell(\gnew)(n,t)=\gnew(n+\ell,t)$.

Since $u_\ik$ corresponds to $u(n+\ik,t)$, we have
\begin{gather}
\lb{csuf}
\cS(u_\ik)=u_{\ik+1},\qquad\cS^\ell(u_\ik)=u_{\ik+\ell},\qquad
\cS^\ell\big(\fn(u_\ik,\dots)\big)=\fn(\cS^\ell(u_{\ik}),\dots)\qquad\quad\forall\,\ell\in\zz.
\end{gather}
Thus, applying $\cS^\ell$ to a function $\fn=\fn(\ud)$, 
one replaces~$u_\ik^{\iup}$ by~$u_{\ik+\ell}^{\iup}$ in~$\fn$ for all $\ik,\iup$.

Using the components $\ff^\iup$ of the vector-function 
$\ff=(\ff^1,\dots,\ff^\diunew)$ from~\eqref{sddenew},
one defines the \emph{total derivative operator}~$\tdt$ acting on
functions of~$u_\ik=(u^1_\ik,\dots,u^\diunew_\ik)$
\begin{gather}
\lb{dtfu}
\tdt\big(\fn(\ud)\big)=\sum_{\ik,\iup}\cS^\ik(\ff^\iup)\cdot\frac{\pd \fn}{\pd u_\ik^\iup}.
\end{gather}
Formula~\er{dtfu} reflects the chain rule for the derivative with respect to~$t$, 
taking into account equations~\er{uildde}.
The presented definitions of~$\cS$ and~$\tdt$ imply
$\tdt\big(\cS(\gnew)\big)=\cS\big(\tdt(\gnew)\big)$ for any function~$\gnew=\gnew(\ud)$.

\begin{definition}
\lb{dmlpgtnew}
Let $\sm\in\zsp$. 
Let $\mm=\mm(\ud,\la)$ and $\lt=\lt(\ud,\la)$ be $\sm\times\sm$ matrices (matrix-functions)
depending on a finite number of the variables~$u_\ik$ and a parameter~$\la$.
Suppose that the matrix~$\mm$ is invertible and $\mm,\,\lt$ satisfy
\begin{gather}
\lb{lr}
\tdt(\mm)=\cS(\lt)\cdot\mm-\mm\cdot\lt.
\end{gather}
Then the pair $(\mm,\,\lt)$ is called a \emph{matrix Lax representation} (MLR) for equation~\eqref{sddenew}.

The connection between~\eqref{lr} and~\eqref{sddenew} is as follows.
The components~$\ff^\iup$ of the right-hand side $\ff(u_\ua,u_{\ua+1},\dots,u_\ub)$ 
of~\eqref{lr} appear in the formula~\er{dtfu} for the operator~$\tdt$, which is used in~\er{lr}.

Relation~\er{lr} implies that the following (overdetermined) auxiliary linear system 
for an invertible $\sm\times\sm$ matrix $\Psi=\Psi(n,t)$
\begin{gather}
\lb{syspsi}
\cS(\Psi)=\mm\cdot\Psi,\qquad\qquad
\pd_t(\Psi)=\lt\cdot\Psi
\end{gather}
is compatible modulo equation~\eqref{sddenew}. 

The matrix~$\mm$ can be called the discrete (difference) part of the MLR.
For brevity, we call~$\mm$ the \emph{$\cS$-part} of the MLR.

Let $(\mm,\,\lt)$ be an MLR for equation~\eqref{sddenew}.
Then for any invertible $\sm\times\sm$ matrix  
$\mathbf{g}=\mathbf{g}(\ud,\la)$ one gets another MLR for~\eqref{sddenew} as follows
\begin{gather}
\lb{ggtmtu}
\hat \mm=\cS(\mathbf{g})\cdot \mm\cdot\mathbf{g}^{-1},\qquad\quad
\hat\lt=\tdt(\mathbf{g})\cdot\mathbf{g}^{-1}+
\mathbf{g}\cdot\lt\cdot\mathbf{g}^{-1}.
\end{gather}
The MLR $(\hat \mm,\,\hat\lt)$ is said to be \emph{gauge equivalent} to the MLR $(\mm,\,\lt)$,
and $\mathbf{g}$ is called a \emph{gauge transformation}.
One can say that $(\hat \mm,\,\hat\lt)$ is derived from $(\mm,\,\lt)$ 
by means of the gauge transformation~$\mathbf{g}$.

Gauge transformations constitute a group with respect to the multiplication of invertible matrices.

Formulas~\er{ggtmtu} define an action of the group of gauge transformations~$\mathbf{g}$
on the set of all MLRs of a given equation~\eqref{sddenew}.
\end{definition}

\begin{remark}
\lb{rdlr}
In~\cite{BIg2016,kmw}, such differential-difference matrix Lax representations
are called Darboux--Lax representations, 
since many of them originate from Darboux transformations of 
(1+1)-dimensional PDEs (see, e.g.,~\cite{kmw}).
\end{remark}

In this paper, for any function $y=y(n,t)$ and each integer~$\ik$ 
we denote by~$y_\ik$ the function of~$n,t$ given by 
$y_\ik(n,t)=y(n+\ik,t)$. In particular, $y_0=y$.


Let $\tua,\tub\in\zz$, $\,\tua\le\tub$. 
Consider another $\diunew$-component evolutionary differential-difference equation
\begin{gather}
\lb{vddenew}
\pd_t(\mv)=\tff(\mv_\tua,\mv_{\tua+1},\dots,\mv_\tub)
\end{gather}
for an $\diunew$-component vector-function $\mv=\big(\mv^1(n,t),\dots,\mv^{\diunew}(n,t)\big)$.

\begin{definition}
\lb{defmtt}
A \emph{Miura-type transformation} (MT) from equation~\eqref{vddenew} 
to equation~\eqref{sddenew} is written as
\begin{gather}
\lb{uvf}
u=\mf(\mv_\ik,\dots)
\end{gather}
with the following requirements:
\begin{enumerate}
	\item the right-hand side of~\er{uvf} 
	is an $\diunew$-component vector-function $\mf=(\mf^1,\dots,\mf^{\diunew})$
	depending on a finite number of the dynamical variables 
	$\mv_\ik=(\mv^1_\ik,\dots,\mv^\diunew_\ik)$, $\ik\in\zz$,
	\item if $\mv=\mv(n,t)$ satisfies equation~\eqref{vddenew} 
	then $u=u(n,t)$ determined by~\eqref{uvf} satisfies equation~\eqref{sddenew}.
\end{enumerate}

More precisely, the second requirement means that we must have relations~\er{dtmfi} explained below.
The components of the vector formula~\eqref{uvf} are 
\begin{gather}
\lb{uimfi}
u^i=\mf^i(\mv^\iup_\ik,\dots),\qquad\quad i=1,\dots,\diunew.
\end{gather}
Substituting the right-hand side of~\eqref{uimfi} in place of~$u^i$ in~\eqref{msdde}, one gets 
\begin{gather}
\lb{dtmfi}
\tdt\big(\mf^i(\mv^\iup_\ik,\dots)\big)=
\ff^i\big(\cS^\ua(\mf^\iup),\cS^{\ua+1}(\mf^\iup),\dots,\cS^\ub(\mf^\iup)\big),\qquad\quad
i=1,\dots,\diunew,
\end{gather}which must be valid identically in the variables $\mv^\iup_\ik$.
\end{definition}

\begin{example}
Let $\diunew=1$. Then $u$ and $\mv$ are scalar functions.
One has the well-known MT $u=\mv_0 \mv_1 \mv_2$ 
from the modified Narita--Itoh--Bogoyavlensky equation 
$\pd_t(\mv)=(\mv_0)^2(\mv_2+\mv_1-\mv_{-1}-\mv_{-2})$
to the Narita--Itoh--Bogoyavlensky equation $\pd_t(u)=u_0(u_2+u_1-u_{-1}-u_{-2})$.
\end{example}

Proposition~\ref{tmlr} is proved in~\cite[Theorem~1]{IgLaxGauge2024}, 
using some ideas from~\cite{IgSimpLax2024}.

\begin{proposition}[\cite{IgLaxGauge2024}]
\label{tmlr}
Let $\diunew,\sm\in\zsp$ and 
$u_{\ik}=(u^1_{\ik},\dots,u^\diunew_{\ik})$ for any $\ik\in\zsp$. 
Consider a $\sm\times\sm$ matrix~$\mm=\mm(u_0,u_1,u_2,\la)$. 
Suppose that 
\begin{gather}
\lb{pdmnew}
\forall\, i,j=1,\dots,\diunew\qquad\quad
\frac{\pd}{\pd u^i_0}\Big(\frac{\pd}{\pd u^j_2}
\big(\mm(u_0,u_1,u_2,\la)\big)\cdot \mm(u_0,u_1,u_2,\la)^{-1}\Big)=0,\\
\lb{pdm01}
\begin{split}
\forall\, i,j&=1,\dots,\diunew\qquad
\frac{\pd}{\pd u^i_0}
\left(\frac{\pd}{\pd u^j_1}\big(\mm(u_0,u_1,u_2,\la)\big)\cdot\mm(u_0,u_1,u_2,\la)^{-1}+\right.\\
&\left.+\mm(u_0,u_1,u_2,\la)\cdot
\frac{\pd}{\pd u^j_1}\big(\mm(a_0,u_0,u_1,\la)\big)\cdot\big(\mm(a_0,u_0,u_1,\la)\big)^{-1}
\cdot\mm(u_0,u_1,u_2,\la)^{-1}\right)=0,
\end{split}
\end{gather}
where $a_0\in\fik^\diunew$ is a constant vector and 
$\mm(a_0,u_0,u_1,\la)=\cS^{-1}\big(\mm(a_0,u_1,u_2,\la)\big)$.

From~\er{pdmnew} it follows that the matrix
\begin{gather}
\lb{tilm}
\tilde{\mm}=\big(\mm(a_0,u_1,u_2,\la)\big)^{-1}\cdot\mm(u_0,u_1,u_2,\la)\cdot\mm(a_0,u_0,u_1,\la)
\end{gather}
does not depend on~$u_2$.
So $\tilde{\mm}$ is of the form $\tilde{\mm}=\tilde{\mm}(u_0,u_{1},\la)$. 

One has the gauge transformation
\begin{gather}
\lb{guu1}
\mathbf{g}(u_0,u_{1},\la)=\big(\tilde{\mm}(\tilde{a}_0,u_0,\la)\big)^{-1}\cdot\big(\mm(a_0,u_0,u_1,\la)\big)^{-1},
\end{gather}
where $\tilde{a}_0\in\fik^\diunew$ is another constant vector and 
$\tilde{\mm}(\tilde{a}_0,u_0,\la)=\cS^{-1}\big(\tilde{\mm}(\tilde{a}_0,u_1,\la)\big)$.

From~\er{pdmnew},~\er{pdm01} it follows that the matrix 
\begin{gather}
\lb{hatmm}
\hat{\mm}=\cS\big(\mathbf{g}(u_0,u_{1},\la)\big)\cdot\mm(u_0,u_1,u_2,\la)\cdot\mathbf{g}(u_0,u_{1},\la)^{-1}
\end{gather}
does not depend on~$u_1,\,u_2$.
So $\hat{\mm}$ is of the form $\hat{\mm}=\hat{\mm}(u_0,\la)$. 

Suppose that one has an MLR $(\mm,\,\lt)$ with~$\mm=\mm(u_0,u_1,u_2,\la)$
satisfying~\er{pdmnew},~\er{pdm01}.
Then, applying the gauge transformation~\er{guu1} to~$(\mm,\,\lt)$, 
we get the gauge equivalent MLR~$(\hat{\mm},\,\hat{\lt})$ given by~\er{hatmm} and
\begin{gather}
\lb{hatlt}
\hat{\lt}=\tdt\big(\mathbf{g}(u_0,u_{1},\la)\big)\cdot\mathbf{g}(u_0,u_{1},\la)^{-1}+
\mathbf{g}(u_0,u_{1},\la)\cdot\lt\cdot\mathbf{g}(u_0,u_{1},\la)^{-1}
\end{gather}
such that $\hat{\mm}$ is of the form $\hat{\mm}=\hat{\mm}(u_0,\la)$.

Thus conditions~\er{pdmnew},~\er{pdm01} are sufficient 
for the possibility to eliminate the dependence on~$u_1,\,u_2$ 
in the~\mbox{$\cS$-part} $\mm=\mm(u_0,u_1,u_2,\la)$
of a given MLR $(\mm,\,\lt)$ by means of a gauge transformation.
\end{proposition}

\section{Integrable equations, Lax representations, 
and Miura-type transformations related to the Belov--Chaltikian lattice}
\lb{sbc}

\subsection{An application of Proposition~\ref{tmlr}}
\lb{sat}
The two-component Belov--Chaltikian lattice, 
which is the Boussinesq lattice related to the lattice 
$W_3$-algebra~\cite{Belov93,hiin97,BeffaWang2013,kmw}, is the equation
\begin{gather}
\label{bce}
\left\{
\begin{aligned}
\pd_t(\bc^1)&= \bc^1 (\bc^2_2-\bc^2_{-1}),\\ 
\pd_t(\bc^2)&= \bc^1_{-1}-\bc^1 + \bc^2 (\bc^2_1-\bc^2_{-1}),
\end{aligned}\right.
\end{gather}
where $\bc^1=\bc^1(n,t)$, $\,\bc^2=\bc^2(n,t)$.
It is known~\cite{hiin97,kmw} that the following matrices form an MLR for~\eqref{bce}
\begin{gather}
\lb{mlrbc}
\mm^{\text{\tiny{BC}}}=\begin{pmatrix} 
\lambda & \lambda \bc^2& \lambda \bc^1_{-1} \\ 
1 & 0& 0\\
0 & 1& 0
\end{pmatrix},\qquad\quad
\lt^{\text{\tiny{BC}}}=  
\begin{pmatrix}
\bc^2-\lambda & -\lambda \bc^2 & -\lambda \bc^1_{-1}\\ 
-1 & \bc^2_{-1} & 0\\
0 & -1 & \bc^2_{-2}
\end{pmatrix}.
\end{gather}

G.~Mar\'i Beffa and Jing~Ping~Wang~\cite{BeffaWang2013} obtained the equation
\begin{gather}
\label{peq}
\left\{
\begin{aligned}
\pd_t(\pe^1)&= \frac{1}{\pe^2_1}-\frac{1}{\pe^2_{-2}},\\ 
\pd_t(\pe^2)&= \frac{\pe^1_1}{\pe^2_1}-\frac{\pe^1}{\pe^2_{-1}}
\end{aligned}\right.
\end{gather}
describing the evolution induced on invariants by an invariant evolution of planar polygons.

It is noticed in~\cite{BeffaWang2013} that the formulas
\begin{gather}
\lb{mtpbc}
\left\{
\begin{aligned}
\bc^1&=
\frac{1}{\pe^2 \pe^2_1 \pe^2_2},\\
\bc^2&=
-\frac{\pe^1_1}{\pe^2 \pe^2_1}\\
\end{aligned}\right.
\end{gather}
determine an MT from~\er{peq} to~\er{bce}.

We relabel 
\begin{gather}
\lb{rbc}
\bc^1_\ik\,\mapsto\,\bc^1_{\ik+1},\qquad\quad \bc^2_\ik\,\mapsto\,\bc^2_\ik\qquad\quad\forall\,\ik\in\zz,\\
\lb{rp}
\pe^1_\ik\,\mapsto\,\pe^1_\ik,\qquad\quad \pe^2_\ik\,\mapsto\,\pe^2_{\ik+1}\qquad\quad\forall\,\ik\in\zz.
\end{gather}
This means that we make the following invertible change of variables 
\begin{gather}
\notag
\bc^1(n,t)\,\mapsto\,\bc^1(n+1,t),\qquad\bc^2(n,t)\,\mapsto\,\bc^2(n,t),\qquad
\pe^1(n,t)\,\mapsto\,\pe^1(n,t),\qquad\pe^2(n,t)\,\mapsto\,\pe^2(n+1,t).
\end{gather}

Applying the change~\er{rbc},~\er{rp} to \er{bce}, \er{mlrbc}, \er{peq}, \er{mtpbc}, we get
\begin{gather}
\label{chbce}
\left\{
\begin{aligned}
\pd_t(\bc^1_1)&= \bc^1_1 (\bc^2_2-\bc^2_{-1}),\\ 
\pd_t(\bc^2)&= \bc^1-\bc^1_1+\bc^2 (\bc^2_1-\bc^2_{-1}),
\end{aligned}\right.\\
\lb{nmlrbc}
\check{\mm}^{\text{\tiny{BC}}}=\begin{pmatrix} 
\lambda & \lambda \bc^2& \lambda \bc^1 \\ 
1 & 0& 0\\
0 & 1& 0
\end{pmatrix},\qquad\quad
\check{\lt}^{\text{\tiny{BC}}}=  
\begin{pmatrix}
\bc^2-\lambda & -\lambda \bc^2 & -\lambda \bc^1\\ 
-1 & \bc^2_{-1} & 0\\
0 & -1 & \bc^2_{-2}
\end{pmatrix},\\
\label{chpeq}
\left\{
\begin{aligned}
\pd_t(\pe^1)&= \frac{1}{\pe^2_2}-\frac{1}{\pe^2_{-1}},\\ 
\pd_t(\pe^2_1)&= \frac{\pe^1_1}{\pe^2_2}-\frac{\pe^1}{\pe^2},
\end{aligned}\right.\\
\lb{chmtpbc}
\left\{
\begin{aligned}
\bc^1_1&=\frac{1}{\pe^2_1 \pe^2_2 \pe^2_3},\\
\bc^2&=-\frac{\pe^1_1}{\pe^2_1 \pe^2_2}.\\
\end{aligned}\right.
\end{gather}
Applying the invertible operator~$\cS^{-1}$ to the first component in~\er{chbce},~\er{chmtpbc}
and the second component in~\er{chpeq}, one obtains
\begin{gather}
\label{nbce}
\left\{
\begin{aligned}
\pd_t(\bc^1)&= \bc^1 (\bc^2_1-\bc^2_{-2}),\\ 
\pd_t(\bc^2)&= \bc^1-\bc^1_1+\bc^2(\bc^2_1-\bc^2_{-1}),
\end{aligned}\right.\\
\label{npeq}
\left\{
\begin{aligned}
\pd_t(\pe^1)&= \frac{1}{\pe^2_2}-\frac{1}{\pe^2_{-1}},\\ 
\pd_t(\pe^2)&= \frac{\pe^1}{\pe^2_1}-\frac{\pe^1_{-1}}{\pe^2_{-1}},
\end{aligned}\right.\\
\lb{nmtpbc}
\left\{
\begin{aligned}
\bc^1&=
\frac{1}{\pe^2_0 \pe^2_1 \pe^2_2},\\
\bc^2&=
-\frac{\pe^1_1}{\pe^2_1 \pe^2_2}.\\
\end{aligned}\right.
\end{gather}
Thus \er{nmlrbc} is an MLR for equation~\er{nbce}, while 
\er{nmtpbc} is an MT from~\er{npeq} to~\er{nbce}.

Below we use the notation~\er{uldiunew} with $\diunew=2$.
That is, for each $\ik\in\zz$ one has $u_\ik=(u^1_\ik,u^2_\ik)$.

Substituting~\er{nmtpbc} in~\er{nmlrbc}, we obtain the following MLR for equation~\er{npeq}
\begin{gather}
\lb{Mbce}
\mm(\pe_{0},\pe_{1},\pe_{2},\la)=
\begin{pmatrix}
 \la & -\frac{\la \pe^1_{1}}{\pe^2_{1} \pe^2_{2}} & \frac{\la}{\pe^2_{0} \pe^2_{1} \pe^2_{2}} \\
 1 & 0 & 0 \\
 0 & 1 & 0 \\
\end{pmatrix},\\
\lb{Ubce}
\lt(\pe_{-1},\pe_{0},\pe_{1},\pe_{2},\la)=
\begin{pmatrix}
 -\frac{\la \pe^2_{1} \pe^2_{2}+\pe^1_{1}}{\pe^2_{1} \pe^2_{2}} & \frac{\la \pe^1_{1}}{\pe^2_{1} \pe^2_{2}} & -\frac{\la}{\pe^2_{0} \pe^2_{1} \pe^2_{2}} \\
 -1 & -\frac{\pe^1_{0}}{\pe^2_{0} \pe^2_{1}} & 0 \\
 0 & -1 & -\frac{\pe^1_{-1}}{\pe^2_{-1} \pe^2_{0}} \\
\end{pmatrix}.
\end{gather}

The matrix~\er{Mbce} satisfies conditions~\er{pdmnew},~\er{pdm01}.
Therefore, we can apply Proposition~\ref{tmlr} to the MLR~\er{Mbce}, \er{Ubce}.
In order to apply Proposition~\ref{tmlr} in this case, we need to
\begin{itemize}
	\item choose a constant vector $a_0\in\fik^2$,
	\item substitute $\pe_0=a_0$ in~$\mm(\pe_{0},\pe_{1},\pe_{2},\la)$
given by~\er{Mbce},
	\item consider the matrix $\mm(a_0,\pe_0,\pe_1,\la)=\cS^{-1}\big(\mm(a_0,\pe_1,\pe_2,\la)\big)$,
	\item compute $\tilde{\mm}=\tilde{\mm}(\pe_0,\pe_{1},\la)$ given by~\er{tilm},
	\item choose another constant vector $\tilde{a}_0\in\fik^2$,
	\item substitute $\pe_0=\tilde{a}_0$ in~$\tilde{\mm}(\pe_0,\pe_{1},\la)$ 
	and consider $\tilde{\mm}(\tilde{a}_0,\pe_0,\la)=\cS^{-1}\big(\tilde{\mm}(\tilde{a}_0,\pe_1,\la)\big)$,
	\item compute $\mathbf{g}(\pe_0,\pe_{1},\la)$ given by~\er{guu1},
	\item apply the obtained gauge transformation $\mathbf{g}(\pe_0,\pe_{1},\la)$ 
	to the MLR~\er{Mbce},~\er{Ubce} and compute the gauge equivalent MLR $(\hat{\mm},\,\hat{\lt})$
	given by~\er{hatmm},~\er{hatlt}.
\end{itemize}
It is convenient to take $a_0=(1,1)$. 
Since $\pe_0=(\pe^1_0,\pe^2_0)$, 
in order to substitute $\pe_0=a_0=(1,1)$ in~$\mm(\pe_{0},\pe_{1},\pe_{2},\la)$, 
we substitute $\pe^1_0=1$, $\,\pe^2_0=1$ in the right-hand side of~\er{Mbce}.
Then 
\begin{gather}
\lb{mu0u1}
\mm(a_0,\pe_0,\pe_1,\la)=\cS^{-1}\big(\mm(a_0,\pe_1,\pe_2,\la)\big)=
\cS^{-1}\begin{pmatrix}
 \la & -\frac{\la \pe^1_{1}}{\pe^2_{1} \pe^2_{2}} & \frac{\la}{\pe^2_{1} \pe^2_{2}} \\
 1 & 0 & 0 \\
 0 & 1 & 0 \\
\end{pmatrix}
=\begin{pmatrix}
 \la & -\frac{\la \pe^1_{0}}{\pe^2_{0} \pe^2_{1}} & \frac{\la}{\pe^2_{0} \pe^2_{1}} \\
 1 & 0 & 0 \\
 0 & 1 & 0 \\
\end{pmatrix}.
\end{gather}
Using~\er{Mbce} and~\er{mu0u1}, 
one computes $\tilde{\mm}=\tilde{\mm}(\pe_0,\pe_{1},\la)$ given by~\er{tilm} as follows
\begin{gather}
\lb{ctilm}
\tilde{\mm}(\pe_0,\pe_{1},\la)=
\big(\mm(a_0,\pe_1,\pe_2,\la)\big)^{-1}\cdot\mm(\pe_0,\pe_1,\pe_2,\la)\cdot\mm(a_0,\pe_0,\pe_1,\la)=
\begin{pmatrix}
 \la & -\frac{\la \pe^1_{0}}{\pe^2_{0} \pe^2_{1}} & \frac{\la}{\pe^2_{0} \pe^2_{1}} \\
 1 & 0 & 0 \\
 0 & \frac{1}{\pe^2_{0}} & 0 \\
\end{pmatrix}.
\end{gather}
Now we take $\tilde{a}_0=(1,1)$ and substitute $\pe_0=\tilde{a}_0=(1,1)$ in~$\tilde{\mm}(\pe_0,\pe_{1},\la)$.
This means that we substitute $\pe^1_0=1$, $\,\pe^2_0=1$ in the right-hand side of~\er{ctilm}.
Then
\begin{gather}
\notag
\tilde{\mm}(\tilde{a}_0,\pe_0,\la)=\cS^{-1}\big(\tilde{\mm}(\tilde{a}_0,\pe_1,\la)\big)=
\cS^{-1}\begin{pmatrix}
 \la & -\frac{\la}{\pe^2_{1}} & \frac{\la}{\pe^2_{1}} \\
 1 & 0 & 0 \\
 0 & 1 & 0 \\
\end{pmatrix}=
\begin{pmatrix}
 \la & -\frac{\la}{\pe^2_{0}} & \frac{\la}{\pe^2_{0}} \\
 1 & 0 & 0 \\
 0 & 1 & 0 \\
\end{pmatrix}.
\end{gather}
This allows us to compute $\mathbf{g}(\pe_0,\pe_{1},\la)$ given by~\er{guu1} as follows
\begin{gather}
\lb{cguu1}
\mathbf{g}(\pe_0,\pe_{1},\la)=
\big(\tilde{\mm}(\tilde{a}_0,\pe_0,\la)\big)^{-1}\cdot\big(\mm(a_0,\pe_0,\pe_1,\la)\big)^{-1}=
\begin{pmatrix}
 0 & 0 & 1 \\
 \frac{\pe^2_{0} \pe^2_{1}}{\la} & -\pe^2_{0} \pe^2_{1} & \pe^1_{0} \\
 \frac{\pe^2_{0} \pe^2_{1}}{\la} & -\frac{\pe^2_{0} (\la \pe^2_{1}-1)}{\la} & \pe^1_{0}-\pe^2_{0} \\
\end{pmatrix}.
\end{gather}
Finally, applying the obtained gauge transformation~\er{cguu1}	to the MLR~\er{Mbce},~\er{Ubce} 
and computing the MLR $(\hat{\mm},\,\hat{\lt})$ given by~\er{hatmm},~\er{hatlt}, we obtain
\begin{gather}
\lb{chmm}
\hat{\mm}(\pe_{0},\la)=
\cS\big(\mathbf{g}(\pe_0,\pe_{1},\la)\big)\cdot\mm(\pe_0,\pe_1,\pe_2,\la)\cdot\mathbf{g}(\pe_0,\pe_{1},\la)^{-1}=
\begin{pmatrix}
 \la & -\frac{\la}{\pe^2_{0}} & \frac{\la}{\pe^2_{0}} \\
 \frac{1}{\pe^2_{0}} & 0 & 0 \\
 -\frac{\pe^1_{0}-1}{\pe^2_{0}} & \frac{1}{\pe^2_{0}} & 0 \\
\end{pmatrix},\\
\lb{chlt}
\begin{split}
\hat{\lt}(\pe_{-1},\pe_{0},\pe_{1},\la)=\tdt\big(\mathbf{g}(\pe_0,\pe_{1},\la)\big)\cdot
&\mathbf{g}(\pe_0,\pe_{1},\la)^{-1}+
\mathbf{g}(\pe_0,\pe_{1},\la)\cdot\lt\cdot\mathbf{g}(\pe_0,\pe_{1},\la)^{-1}=\\
&=
\begin{pmatrix}
 -\frac{\la \pe^2_{-1} \pe^2_{0}+\pe^1_{-1}}{\pe^2_{-1} \pe^2_{0}} & \frac{\la}{\pe^2_{0}} & -\frac{\la}{\pe^2_{0}} \\
 -\frac{1}{\pe^2_{-1}} & -\frac{\pe^1_{-1}}{\pe^2_{-1} \pe^2_{0}} & 0 \\
 \frac{\pe^1_{-1}-1}{\pe^2_{-1}} & -\frac{1}{\pe^2_{1}} & -\frac{\pe^1_{-1}}{\pe^2_{-1} \pe^2_{0}} \\
\end{pmatrix}.
\end{split}
\end{gather}
In agreement with Proposition~\ref{tmlr}, we see that
the matrix~\er{chmm} depends only on~$\pe_{0},\,\la$,
in contrast to the matrix~\er{Mbce} depending on~$\pe_{0},\,\pe_{1},\,\pe_2,\,\la$.

\subsection{Constructions of new Miura-type transformations and integrable equations}
\lb{scmt}

Fix a constant $\ap\in\mathbb{C}$.
Below we consider the matrices~\er{chmm},~\er{chlt} with~$\la$ replaced by~$\ap$.

According to Definition~\ref{dmlpgtnew}, since the matrices~\er{chmm},~\er{chlt} 
form an MLR for equation~\er{npeq}, we can consider the auxiliary linear system
\begin{gather}
\lb{syspsi3}
\cS(\Psi)=\mm\cdot\Psi,\qquad\qquad\pd_t(\Psi)=\lt\cdot\Psi,
\end{gather}
which is compatible modulo equation~\eqref{npeq}. 
Here $\Psi=\Psi(n,t)$ is an invertible $3\times 3$ matrix
whose elements we denote by $\uppsi^{ij}=\uppsi^{ij}(n,t)$, $\,i,j=1,2,3$.

Let 
$$
\hat{\mme}^{ij}=\hat{\mme}^{ij}(\pe_{0},\ap),\qquad\quad
\hat{\lte}^{ij}=\hat{\lte}^{ij}(\pe_{-1},\pe_0,\pe_1,\ap),\qquad\quad i,j=1,2,3,
$$
be the elements of the $3\times 3$ matrices $\hat{\mm}$, $\,\hat{\lt}$ given by~\er{chmm},~\er{chlt} 
with~$\la$ replaced by~$\ap$. We have
\begin{gather}
\label{echmm}
\hat{\mm}(\pe_{0},\ap)=
\begin{pmatrix}
 \ap & -\frac{\ap}{\pe^2_{0}} & \frac{\ap}{\pe^2_{0}} \\
 \frac{1}{\pe^2_{0}} & 0 & 0 \\
 -\frac{\pe^1_{0}-1}{\pe^2_{0}} & \frac{1}{\pe^2_{0}} & 0 \\
\end{pmatrix}=
\begin{pmatrix}
\hat{\mme}^{11}(\pe_{0},\ap) & \hat{\mme}^{12}(\pe_{0},\ap) & \hat{\mme}^{13}(\pe_{0},\ap) \\
\hat{\mme}^{21}(\pe_{0},\ap) & \hat{\mme}^{22}(\pe_{0},\ap) & \hat{\mme}^{23}(\pe_{0},\ap) \\
\hat{\mme}^{31}(\pe_{0},\ap) & \hat{\mme}^{32}(\pe_{0},\ap) & \hat{\mme}^{33}(\pe_{0},\ap)
\end{pmatrix},\\
\label{echlt}
\begin{split}
\hat{\lt}(\pe_{-1},\pe_0,\pe_1,\ap)&=
\begin{pmatrix}
 -\frac{\ap \pe^2_{-1} \pe^2_{0}+\pe^1_{-1}}{\pe^2_{-1} \pe^2_{0}} & \frac{\ap}{\pe^2_{0}} & -\frac{\ap}{\pe^2_{0}} \\
 -\frac{1}{\pe^2_{-1}} & -\frac{\pe^1_{-1}}{\pe^2_{-1} \pe^2_{0}} & 0 \\
 \frac{\pe^1_{-1}-1}{\pe^2_{-1}} & -\frac{1}{\pe^2_{1}} & -\frac{\pe^1_{-1}}{\pe^2_{-1} \pe^2_{0}} \\
\end{pmatrix}=\\
&=
\begin{pmatrix}
\hat{\lte}^{11}(\pe_{-1},\pe_0,\pe_1,\ap) & \hat{\lte}^{12}(\pe_{-1},\pe_0,\pe_1,\ap) & \hat{\lte}^{13}(\pe_{-1},\pe_0,\pe_1,\ap) \\
\hat{\lte}^{21}(\pe_{-1},\pe_0,\pe_1,\ap) & \hat{\lte}^{22}(\pe_{-1},\pe_0,\pe_1,\ap) & \hat{\lte}^{23}(\pe_{-1},\pe_0,\pe_1,\ap) \\
\hat{\lte}^{31}(\pe_{-1},\pe_0,\pe_1,\ap) & \hat{\lte}^{32}(\pe_{-1},\pe_0,\pe_1,\ap) & \hat{\lte}^{33}(\pe_{-1},\pe_0,\pe_1,\ap)
\end{pmatrix}.
\end{split}
\end{gather}
System~\er{syspsi3} reads
\begin{gather}
\label{psi11}
\begin{aligned}
\cS\begin{pmatrix}
\uppsi^{11}&\uppsi^{12}&\uppsi^{13}\\
\uppsi^{21}&\uppsi^{22}&\uppsi^{23}\\
\uppsi^{31}&\uppsi^{32}&\uppsi^{33}
\end{pmatrix}&=
\begin{pmatrix}
\hat{\mme}^{11} & \hat{\mme}^{12} & \hat{\mme}^{13} \\
\hat{\mme}^{21} & \hat{\mme}^{22} & \hat{\mme}^{23} \\
\hat{\mme}^{31} & \hat{\mme}^{32} & \hat{\mme}^{33}
\end{pmatrix}
\begin{pmatrix}
\uppsi^{11}&\uppsi^{12}&\uppsi^{13}\\
\uppsi^{21}&\uppsi^{22}&\uppsi^{23}\\
\uppsi^{31}&\uppsi^{32}&\uppsi^{33}
\end{pmatrix},\\
\pd_t\begin{pmatrix}
\uppsi^{11}&\uppsi^{12}&\uppsi^{13}\\
\uppsi^{21}&\uppsi^{22}&\uppsi^{23}\\
\uppsi^{31}&\uppsi^{32}&\uppsi^{33}
\end{pmatrix}&=\begin{pmatrix}
\hat{\lte}^{11} & \hat{\lte}^{12} & \hat{\lte}^{13} \\
\hat{\lte}^{21} & \hat{\lte}^{22} & \hat{\lte}^{23} \\
\hat{\lte}^{31} & \hat{\lte}^{32} & \hat{\lte}^{33}
\end{pmatrix}
\begin{pmatrix}
\uppsi^{11}&\uppsi^{12}&\uppsi^{13}\\
\uppsi^{21}&\uppsi^{22}&\uppsi^{23}\\
\uppsi^{31}&\uppsi^{32}&\uppsi^{33}
\end{pmatrix}.
\end{aligned}
\end{gather}
From~\er{psi11} one gets
\begin{gather}
\lb{spdp}
\begin{aligned}
\cS(\uppsi^{ij})&=\hat{\mme}^{i1}\uppsi^{1j}+\hat{\mme}^{i2}\uppsi^{2j}+\hat{\mme}^{i3}\uppsi^{3j},\\
\pd_t(\uppsi^{ij})&=\hat{\lte}^{i1}\uppsi^{1j}+\hat{\lte}^{i2}\uppsi^{2j}+\hat{\lte}^{i3}\uppsi^{3j},\qquad\quad
i,j=1,2,3.
\end{aligned}
\end{gather}
We set $\vv^1=\dfrac{\uppsi^{11}}{\uppsi^{21}}$ and $\vv^2=\dfrac{\uppsi^{31}}{\uppsi^{21}}$.
From~\er{spdp} it follows that
\begin{gather}
\label{svv1}
\cS(\vv^1)=\frac{\cS(\uppsi^{11})}{\cS(\uppsi^{21})}=
\frac{\hat{\mme}^{11}\uppsi^{11}+\hat{\mme}^{12}\uppsi^{21}+\hat{\mme}^{13}\uppsi^{31}}{\hat{\mme}^{21}\uppsi^{11}+\hat{\mme}^{22}\uppsi^{21}+\hat{\mme}^{23}\uppsi^{31}}=
\frac{\dfrac{\hat{\mme}^{11}\uppsi^{11}}{\uppsi^{21}}+\hat{\mme}^{12}+\dfrac{\hat{\mme}^{13}\uppsi^{31}}{\uppsi^{21}}}{\dfrac{\hat{\mme}^{21}\uppsi^{11}}{\uppsi^{21}}+\hat{\mme}^{22}+\dfrac{\hat{\mme}^{23}\uppsi^{31}}{\uppsi^{21}}}=
\frac{\hat{\mme}^{11}\vv^1+\hat{\mme}^{12}+\hat{\mme}^{13}\vv^2}{\hat{\mme}^{21}\vv^1+\hat{\mme}^{22}+\hat{\mme}^{23}\vv^2},\\
\label{svv2}
\cS(\vv^2)=\frac{\cS(\uppsi^{31})}{\cS(\uppsi^{21})}=
\frac{\hat{\mme}^{31}\uppsi^{11}+\hat{\mme}^{32}\uppsi^{21}+\hat{\mme}^{33}\uppsi^{31}}{\hat{\mme}^{21}\uppsi^{11}+\hat{\mme}^{22}\uppsi^{21}+\hat{\mme}^{23}\uppsi^{31}}=
\frac{\dfrac{\hat{\mme}^{31}\uppsi^{11}}{\uppsi^{21}}+\hat{\mme}^{32}+\dfrac{\hat{\mme}^{33}\uppsi^{31}}{\uppsi^{21}}}{\dfrac{\hat{\mme}^{21}\uppsi^{11}}{\uppsi^{21}}+\hat{\mme}^{22}+\dfrac{\hat{\mme}^{23}\uppsi^{31}}{\uppsi^{21}}}=
\frac{\hat{\mme}^{31}\vv^1+\hat{\mme}^{32}+\hat{\mme}^{33}\vv^2}{\hat{\mme}^{21}\vv^1+\hat{\mme}^{22}+\hat{\mme}^{23}\vv^2},\\
\label{dvv1}
\begin{split}
\pd_t(\vv^1)&=\pd_t\Big(\frac{\uppsi^{11}}{\uppsi^{21}}\Big)=
\frac{\pd_t(\uppsi^{11})}{\uppsi^{21}}-\frac{\uppsi^{11}\pd_t(\uppsi^{21})}{(\uppsi^{21})^2}=\\
&=\frac{\hat{\lte}^{11}\uppsi^{11}+\hat{\lte}^{12}\uppsi^{21}+\hat{\lte}^{13}\uppsi^{31}}{\uppsi^{21}}-
\frac{\uppsi^{11}\big(\hat{\lte}^{21}\uppsi^{11}+\hat{\lte}^{22}\uppsi^{21}+\hat{\lte}^{23}\uppsi^{31}\big)}{(\uppsi^{21})^2}=\\
&=\hat{\lte}^{11}\frac{\uppsi^{11}}{\uppsi^{21}}+\hat{\lte}^{12}+\hat{\lte}^{13}\frac{\uppsi^{31}}{\uppsi^{21}}
-\hat{\lte}^{21}\Big(\frac{\uppsi^{11}}{\uppsi^{21}}\Big)^2-
\hat{\lte}^{22}\frac{\uppsi^{11}}{\uppsi^{21}}-\hat{\lte}^{23}\frac{\uppsi^{11}}{\uppsi^{21}}\frac{\uppsi^{31}}{\uppsi^{21}}=\\
&=-\hat{\lte}^{21}(\vv^1)^2-\hat{\lte}^{23}\vv^1\vv^2+(\hat{\lte}^{11}-\hat{\lte}^{22})\vv^1+\hat{\lte}^{13}\vv^2+
\hat{\lte}^{12},
\end{split}\\
\label{dvv2}
\begin{split}
\pd_t(\vv^2)&=\pd_t\Big(\frac{\uppsi^{31}}{\uppsi^{21}}\Big)=
\frac{\pd_t(\uppsi^{31})}{\uppsi^{21}}-\frac{\uppsi^{31}\pd_t(\uppsi^{21})}{(\uppsi^{21})^2}=\\
&=\frac{\hat{\lte}^{31}\uppsi^{11}+\hat{\lte}^{32}\uppsi^{21}+\hat{\lte}^{33}\uppsi^{31}}{\uppsi^{21}}-
\frac{\uppsi^{31}\big(\hat{\lte}^{21}\uppsi^{11}+\hat{\lte}^{22}\uppsi^{21}+\hat{\lte}^{23}\uppsi^{31}\big)}{(\uppsi^{21})^2}=\\
&=\hat{\lte}^{31}\frac{\uppsi^{11}}{\uppsi^{21}}+\hat{\lte}^{32}+\hat{\lte}^{33}\frac{\uppsi^{31}}{\uppsi^{21}}
-\hat{\lte}^{21}\frac{\uppsi^{31}}{\uppsi^{21}}\frac{\uppsi^{11}}{\uppsi^{21}}-
\hat{\lte}^{22}\frac{\uppsi^{31}}{\uppsi^{21}}-\hat{\lte}^{23}\Big(\frac{\uppsi^{31}}{\uppsi^{21}}\Big)^2=\\
&=-\hat{\lte}^{21}\vv^2\vv^1-\hat{\lte}^{23}(\vv^2)^2+\hat{\lte}^{31}\vv^1+(\hat{\lte}^{33}-\hat{\lte}^{22})\vv^2
+\hat{\lte}^{32}.
\end{split}
\end{gather}
Equation~\er{echmm} gives us explicit formulas for the functions
$\hat{\mme}^{ij}=\hat{\mme}^{ij}(\pe^1_{0},\pe^2_{0},\ap)$, $\,i,j=1,2,3$.
Substituting these formulas in~\er{svv1},~\er{svv2}, one obtains
\begin{gather}
\label{sv1e}
\cS(\vv^1)=\frac{\ap (\vv^1_{0} \pe^2_{0}+\vv^2_{0}-1)}{\vv^1_{0}},\\
\label{sv2e}
\cS(\vv^2)=-\frac{\vv^1_{0} \pe^1_{0}-\vv^1_{0}-1}{\vv^1_{0}}.
\end{gather}
Below we use the notation 
\begin{gather}
\notag
v^1_\ik=\cS^\ik(v^1),\qquad\quad
v^2_\ik=\cS^\ik(v^2)\qquad\quad\forall\,\ik\in\zz.
\end{gather}
From~\er{sv1e},~\er{sv2e} one gets
\begin{gather}
\lb{mtpe}
\left\{
\begin{aligned}
\pe^1_0&=-\vv^2_{1}+\frac{1}{\vv^1_{0}}+1,\\
\pe^2_0&=\frac{1}{\ap}\vv^1_{1}-\frac{\vv^2_{0}}{\vv^1_{0}}+\frac{1}{\vv^1_{0}},\\
\end{aligned}\right.
\end{gather}
which implies
\begin{gather}
\lb{pe12ell}
\forall\,\ik\in\zz\qquad\quad
\pe^1_\ik=-\vv^2_{\ik+1}+\frac{1}{\vv^1_{\ik}}+1,\qquad\quad
\pe^2_\ik=\frac{1}{\ap}\vv^1_{\ik+1}-\frac{\vv^2_{\ik}}{\vv^1_{\ik}}
+\frac{1}{\vv^1_{\ik}}.
\end{gather}

Equation~\er{echlt} yields explicit formulas for the functions
$$
\hat{\lte}^{ij}=\hat{\lte}^{ij}(\pe^1_{-1},\pe^2_{-1},\pe^1_0,\pe^2_0,\pe^1_1,\pe^2_1,\ap),
\qquad\quad i,j=1,2,3.
$$
We substitute these formulas along with~\er{pe12ell} to equations~\er{dvv1},~\er{dvv2}, 
which gives 
\begin{gather}
\label{dtv}
\left\{
\begin{aligned}
\pd_t(\vv^1)&=-\frac{(\ap\vv^1_{0})^2 (\vv^2_{0} \vv^1_{-1}-\vv^1_{-1}+\vv^1_{1}-\vv^1_{1} \vv^2_{-1})}{(-\ap \vv^2_{-1}+\vv^1_{-1} \vv^1_{0}+\ap) (-\ap \vv^2_{0}+\vv^1_{0} \vv^1_{1}+\ap)},\\
\pd_t(\vv^2)&=\frac{\ap (-\ap \vv^2_{-1} \vv^1_{1}+\ap \vv^1_{1}-\ap \vv^1_{0}+\ap \vv^1_{0} \vv^2_{1}+\vv^1_{-1} \vv^1_{0} \vv^1_{1}-\vv^1_{0} \vv^1_{2} \vv^1_{1})}{(-\ap \vv^2_{-1}+\vv^1_{-1} \vv^1_{0}+\ap) (\ap \vv^2_{1}-\vv^1_{1} \vv^1_{2}-\ap)}.
\end{aligned}\right.
\end{gather}

Recall that system~\er{syspsi3} is compatible modulo equation~\eqref{npeq}
and is equivalent to~\er{spdp}. 
Since equations~\er{mtpe},~\er{dtv} are derived from~\er{spdp}, 
system~\er{mtpe},~\er{dtv} is compatible modulo equation~\eqref{npeq} as well.
This implies that \er{mtpe} is an MT from~\er{dtv} to~\eqref{npeq}.

Thus \er{mtpe} is an MT for equation~\er{npeq}.
As explained above, \er{npeq} is obtained from~\er{peq} 
by means of an invertible transformation involving~\er{rp}.
In order to return from~\er{npeq} to~\er{peq} and to derive an MT for equation~\eqref{peq},
we are going to use the relabeling
\begin{gather}
\lb{invrp}
\pe^1_\ik\,\mapsto\,\pe^1_\ik,\qquad\pe^2_\ik\,\mapsto\,\pe^2_{\ik-1}\qquad\quad\forall\,\ik\in\zz,
\end{gather}
which is the inverse of~\er{rp}. 
That is, we make the change $\pe^1(n,t)\mapsto\pe^1(n,t)$, 
$\,\pe^2(n,t)\mapsto\pe^2(n-1,t)$.

Applying the relabeling~\er{invrp} to~\er{npeq},~\er{mtpe},~\er{chmm},~\er{chlt}, we get
\begin{gather}
\label{inpeq}
\left\{
\begin{aligned}
\pd_t(\pe^1)&= \frac{1}{\pe^2_1}-\frac{1}{\pe^2_{-2}},\\ 
\pd_t(\pe^2_{-1})&= \frac{\pe^1}{\pe^2}-\frac{\pe^1_{-1}}{\pe^2_{-2}},
\end{aligned}\right.\\
\lb{imtpe}
\left\{
\begin{aligned}
\pe^1_0&=-\vv^2_{1}+\frac{1}{\vv^1_{0}}+1,\\
\pe^2_{-1}&=\frac{1}{\ap}\vv^1_{1}-\frac{\vv^2_{0}}{\vv^1_{0}}+\frac{1}{\vv^1_{0}},\\
\end{aligned}\right.\\
\lb{rchmu}
\check{\mm}=
\begin{pmatrix}
 \la & -\frac{\la}{\pe^2_{-1}} & \frac{\la}{\pe^2_{-1}} \\
 \frac{1}{\pe^2_{-1}} & 0 & 0 \\
 -\frac{\pe^1_{0}-1}{\pe^2_{-1}} & \frac{1}{\pe^2_{-1}} & 0 \\
\end{pmatrix},\qquad\quad
\check{\lt}=
\begin{pmatrix}
 -\frac{\la \pe^2_{-2} \pe^2_{-1}+\pe^1_{-1}}{\pe^2_{-2} \pe^2_{-1}} & \frac{\la}{\pe^2_{-1}} & -\frac{\la}{\pe^2_{-1}} \\
 -\frac{1}{\pe^2_{-2}} & -\frac{\pe^1_{-1}}{\pe^2_{-2} \pe^2_{-1}} & 0 \\
 \frac{\pe^1_{-1}-1}{\pe^2_{-2}} & -\frac{1}{\pe^2_{0}} & -\frac{\pe^1_{-1}}{\pe^2_{-2} \pe^2_{-1}} \\
\end{pmatrix}.
\end{gather}
Applying the operator~$\cS$ to the second component in~\er{inpeq} and~\er{imtpe}, 
one obtains~\er{peq} and 
\begin{gather}
\lb{mtpeqq}
\left\{
\begin{aligned}
\pe^1_0&=-\vv^2_{1}+\frac{1}{\vv^1_{0}}+1,\\
\pe^2_0&=\frac{1}{\ap}\vv^1_{2}-\frac{\vv^2_{1}}{\vv^1_{1}}+\frac{1}{\vv^1_{1}}.\\
\end{aligned}\right.
\end{gather}
Thus \er{mtpeqq} is an MT from~\er{dtv} to~\er{peq}, 
while \er{rchmu} is an MLR for~\er{peq}.

Computing the composition of the MTs~\er{mtpbc} and~\er{mtpeqq}, 
we get the following MT
\begin{gather}
\lb{cmts}
\left\{
\begin{aligned}
\bc^1&=\frac{\ap^3 \vv^1_{1} \vv^1_{2} \vv^1_{3}}{(\ap \vv^2_{1}-\vv^1_{1} \vv^1_{2}-\ap) (\vv^1_{2} \vv^1_{3}-\ap \vv^2_{2}+\ap) (\ap \vv^2_{3}-\vv^1_{3} \vv^1_{4}-\ap)},\\
\bc^2&=\frac{\ap^2 \vv^1_{2} (\vv^2_{2} \vv^1_{1}-\vv^1_{1}-1)}{(\ap \vv^2_{1}-\vv^1_{1} \vv^1_{2}-\ap) (\ap \vv^2_{2}-\vv^1_{2} \vv^1_{3}-\ap)}\\
\end{aligned}\right.
\end{gather}
from equation~\er{dtv} to the Belov--Chaltikian lattice~\er{bce}.

\begin{remark}
\lb{rintdtv}
As shown above, the matrices~\er{chmm},~\er{chlt} form a $\la$-dependent MLR for~\er{npeq}, 
while formula~\er{mtpe} is an MT from~\er{dtv} to~\eqref{npeq}.
It is easily seen that these two facts imply the following.
Substituting~\er{mtpe} in~\er{chmm},~\er{chlt}, one 
obtains a $\la$-dependent MLR for equation~\er{dtv}.  
This MLR for~\er{dtv} seems to be new.

Equation~\er{dtv} is integrable in the sense that 
\er{dtv} possesses a nontrivial MLR with parameter~$\la$
and is connected by the MT~\er{cmts} 
to the well-known integrable Belov--Chaltikian lattice equation~\er{bce}.
\end{remark}

\section{Integrable equations, Lax representations, 
and Miura-type transformations related to the Toda lattice}
\lb{stoda}

Consider the (non-evolutionary) Toda lattice equation for a scalar function $\upbeta=\upbeta(n,t)$
\begin{equation}
\lb{todann}
\frac{\pd^2\upbeta}{\pd t^2}=\exp(\upbeta_{1}-\upbeta)-\exp(\upbeta-\upbeta_{-1}),\qquad\quad
\upbeta_1=\upbeta(n+1,t),\qquad \upbeta_{-1}=\upbeta(n-1,t).
\end{equation}
Following H.~Flaschka and S.V.~Manakov~\cite{flaschka74,manakov75}, 
consider the functions 
$$
\tdnew^1(n,t)=\exp(\upbeta-\upbeta_{-1}),\qquad\quad\tdnew^2(n,t)=\pd_t(\upbeta).
$$
Then equation~\eqref{todann} yields the two-component evolutionary equation
\begin{equation}
\label{todanew}
\left\{
\begin{aligned}
\tdnew^1_t&=\tdnew^1(\tdnew^2-\tdnew^2_{-1}),\\
\tdnew^2_t&=\tdnew^1_1-\tdnew^1,
\end{aligned}\right.
\end{equation}
which is sometimes called the \emph{Toda lattice in the Flaschka--Manakov coordinates}.

It is known that the following matrices form an MLR for~\eqref{todanew}
\begin{equation}
\lb{lrnew}
\mm=\begin{pmatrix} 
\lambda+\tdnew^2 & \tdnew^1 \\ -1 & 0 
\end{pmatrix},\qquad\quad
\lt=  
\begin{pmatrix}
0 &-\tdnew^1\\
1&\lambda+\tdnew^2_{-1}
\end{pmatrix}.
\end{equation}

The following result is presented in~\cite{IgLaxGauge2024}.
Fix constants $\cto,\ctt\in\mathbb{C}$. The two-component equation
\begin{gather}
\lb{mtoda}
\left\{
\begin{gathered}
u^1_t=\frac{u^1_{0}\cdot 
Z^1(u^{1}_{-1},u^{2}_{-1},u^{1}_{0},u^{2}_{0},u^{1}_{1},u^{2}_{1},\cto,\ctt)}{(u^1_{-1}-u^2_{-1}) (u^1_{0}-u^2_{0})},\\
\begin{split}
Z^1&(u^{1}_{-1},u^{2}_{-1},u^{1}_{0},u^{2}_{0},u^{1}_{1},u^{2}_{1},\cto,\ctt)=
\cto u^2_{-1} u^1_{0}-\ctt u^2_{-1} u^1_{0}-\cto u^1_{-1} u^2_{0}+\ctt u^1_{-1} u^2_{0}+\\
&+u^2_{-1} (u^1_{0})^2-2 u^2_{-1} u^2_{0} u^1_{0}
+u^2_{-1} (u^2_{0})^2-u^1_{-1} u^1_{1} u^2_{0}+u^1_{1} u^2_{-1} u^2_{0}+u^1_{-1} u^2_{0} u^2_{1}-u^2_{-1} u^2_{0} u^2_{1},
\end{split}\\
u^2_t=-\frac{u^2_{0} \cdot 
Z^2(u^{1}_{-1},u^{2}_{-1},u^{1}_{0},u^{2}_{0},u^{1}_{1},u^{2}_{1},\cto,\ctt)}{(u^1_{-1}-u^2_{-1}) (u^2_{0}-u^1_{0})},\\
\begin{split}
Z^2&(u^{1}_{-1},u^{2}_{-1},u^{1}_{0},u^{2}_{0},u^{1}_{1},u^{2}_{1},\cto,\ctt)=\cto u^2_{-1} u^1_{0}-\ctt u^2_{-1} u^1_{0}-\cto u^1_{-1} u^2_{0}+\ctt u^1_{-1} u^2_{0}+\\
&+u^1_{-1} (u^1_{0})^2-u^1_{-1} u^1_{1} u^1_{0}+u^1_{1} u^2_{-1} u^1_{0}-2 u^1_{-1} u^2_{0} u^1_{0}+u^1_{-1} u^2_{1} u^1_{0}-u^2_{-1} u^2_{1} u^1_{0}+u^1_{-1} (u^2_{0})^2,
\end{split}
\end{gathered}\right.
\end{gather}
is connected to~\er{todanew} by the MT
\begin{gather}
\lb{dstoda}
\left\{
\begin{aligned}
\tdnew^1&=
\frac{u^1_{0} u^2_{0} (u^1_{1}-u^2_{1}+\cto-\ctt)}{u^1_{0}-u^2_{0}},\\
\tdnew^2&=
-\frac{\cto u^1_{0}-\ctt u^2_{0}+u^1_{1} u^1_{0}-u^2_{0} u^2_{1}}{u^1_{0}-u^2_{0}}.\\
\end{aligned}\right.
\end{gather}


Below we use the notation~\er{uldiunew} with $\diunew=2$.
That is, for each $\ik\in\zz$ one has $u_\ik=(u^1_\ik,u^2_\ik)$.
As shown in~\cite{IgLaxGauge2024}, 
substituting~\er{dstoda} in~\er{lrnew}, one obtains the following MLR for equation~\er{mtoda}
\begin{gather}
\lb{Mmtoda}
\mm(u_{0},u_{1},\la)=\begin{pmatrix} 
\lambda-\frac{\cto u^1_{0}-\ctt u^2_{0}+u^1_{1} u^1_{0}-u^2_{0} u^2_{1}}{u^1_{0}-u^2_{0}} & 
\frac{u^1_{0} u^2_{0} (u^1_{1}-u^2_{1}+\cto-\ctt)}{u^1_{0}-u^2_{0}} \\ 
-1 & 0 
\end{pmatrix},\\
\lb{Umtoda}
\lt(u_{-1},u_{0},u_{1},\la)=\begin{pmatrix}
0 &-\frac{u^1_{0} u^2_{0} (u^1_{1}-u^2_{1}+\cto-\ctt)}{u^1_{0}-u^2_{0}}\\
1&\lambda
-\frac{\cto u^1_{-1}-\ctt u^2_{-1}+u^1_{0} u^1_{-1}-u^2_{-1} u^2_{0}}{u^1_{-1}-u^2_{-1}}
\end{pmatrix}.
\end{gather}

As noticed in~\cite{IgLaxGauge2024}, 
the matrix~\er{Mmtoda} satisfies conditions~\er{pdmnew},~\er{pdm01}, 
hence one can use Proposition~\ref{tmlr}.
(Note that Proposition~\ref{tmlr} does not require nontrivial dependence on~$u_2$
in the matrix~$\mm$.)
Applying Proposition~\ref{tmlr} to the MLR~\er{Mmtoda},~\er{Umtoda},
we get the gauge equivalent MLR
\begin{gather}
\lb{hmhlt}
\hat{\mm}(u_{0},\la)=\cS\big(\mathbf{g}\big)\cdot\mm(u_0,u_1,\la)\cdot\mathbf{g}^{-1},\qquad
\hat{\lt}(u_{-1},u_0,\la)=\tdt(\mathbf{g})\cdot\mathbf{g}^{-1}+
\mathbf{g}\cdot\lt(u_{-1},u_{0},u_{1},\la)\cdot\mathbf{g}^{-1},
\end{gather}
where the gauge transformation~$\mathbf{g}$ is given by~\er{guu1} 
and $\tdt$ is the total derivative operator corresponding to equation~\eqref{mtoda}.
However, in this case the straightforward application of Proposition~\ref{tmlr} 
gives rather cumbersome formulas.

Below we use a different approach, in order to find a gauge 
transformation~$\breve{\mathbf{g}}=\breve{\mathbf{g}}(u_{0},\la)$ such that 
in the corresponding gauge equivalent MLR
\begin{gather}
\label{hmgu1}
\breve{\mm}=
\cS\big(\breve{\mathbf{g}}(u_0,\la)\big)\cdot\mm(u_0,u_1,\la)\cdot\breve{\mathbf{g}}(u_0,\la)^{-1},\\
\label{hugu1}
\breve{\lt}=\tdt\big(\breve{\mathbf{g}}(u_0,\la)\big)\cdot\breve{\mathbf{g}}(u_0,\la)^{-1}+
\breve{\mathbf{g}}(u_0,\la)\cdot\lt(u_{-1},u_{0},u_{1},\la)\cdot\breve{\mathbf{g}}(u_0,\la)^{-1}
\end{gather}
the $\cS$-part~$\breve{\mm}$ depends only on~$u_{0},\la$.

We want to find an invertible matrix~$\breve{\mathbf{g}}(u_0,\la)$ such that
the right-hand side of~\er{hmgu1} does not depend on~$u_1=(u^1_1,u^2_1)$. 
That is, we want to find~$\breve{\mathbf{g}}(u_0,\la)$ such that
\begin{gather}
\label{pdu11}
\frac{\pd}{\pd u^1_1}
\Big(\cS\big(\breve{\mathbf{g}}(u_0,\la)\big)\cdot \mm(u_{0},u_{1},\la)\cdot\breve{\mathbf{g}}(u_0,\la)^{-1}\Big)=0,\\
\label{pdu21}
\frac{\pd}{\pd u^2_1}
\Big(\cS\big(\breve{\mathbf{g}}(u_0,\la)\big)\cdot \mm(u_{0},u_{1},\la)\cdot\breve{\mathbf{g}}(u_0,\la)^{-1}\Big)=0.
\end{gather}
Since $\breve{\mathbf{g}}(u_0,\la)^{-1}$ does not depend on~$u_1=(u^1_1,u^2_1)$
and is required to be invertible,
equations~\er{pdu11},~\er{pdu21} are equivalent to
\begin{gather}
\label{pdu11s}
\frac{\pd}{\pd u^1_1}
\Big(\cS\big(\breve{\mathbf{g}}(u_0,\la)\big)\cdot \mm(u_{0},u_{1},\la)\Big)=0,\\
\label{pdu21s}
\frac{\pd}{\pd u^2_1}
\Big(\cS\big(\breve{\mathbf{g}}(u_0,\la)\big)\cdot \mm(u_{0},u_{1},\la)\Big)=0,
\end{gather}
which say that
\begin{gather}
\label{pdu11p}
\frac{\pd}{\pd u^1_1}\Big(\cS\big(\breve{\mathbf{g}}(u_0,\la)\big)\Big)\cdot \mm(u_{0},u_{1},\la)+
\cS\big(\breve{\mathbf{g}}(u_0,\la)\big)\cdot\frac{\pd}{\pd u^1_1}\big(\mm(u_{0},u_{1},\la)\big)=0,\\
\label{pdu21p}
\frac{\pd}{\pd u^2_1}\Big(\cS\big(\breve{\mathbf{g}}(u_0,\la)\big)\Big)\cdot \mm(u_{0},u_{1},\la)+
\cS\big(\breve{\mathbf{g}}(u_0,\la)\big)\cdot\frac{\pd}{\pd u^2_1}\big(\mm(u_{0},u_{1},\la)\big)=0.
\end{gather}
Multiplying~\er{pdu11p},~\er{pdu21p} by~$\mm(u_{0},u_{1},\la)^{-1}$ from the right, 
we can rewrite~\er{pdu11p},~\er{pdu21p} as
\begin{gather}
\label{pdu11e}
\frac{\pd}{\pd u^1_1}\Big(\cS\big(\breve{\mathbf{g}}(u_0,\la)\big)\Big)=
-\cS\big(\breve{\mathbf{g}}(u_0,\la)\big)\cdot\frac{\pd}{\pd u^1_1}\big(\mm(u_{0},u_{1},\la)\big)
\cdot \mm(u_{0},u_{1},\la)^{-1},\\
\label{pdu21e}
\frac{\pd}{\pd u^2_1}\Big(\cS\big(\breve{\mathbf{g}}(u_0,\la)\big)\Big)=
-\cS\big(\breve{\mathbf{g}}(u_0,\la)\big)\cdot\frac{\pd}{\pd u^2_1}\big(\mm(u_{0},u_{1},\la)\big)
\cdot \mm(u_{0},u_{1},\la)^{-1}.
\end{gather}
Applying the operator~$\cS^{-1}$ to~\er{pdu11e},~\er{pdu21e}, we get the differential equations
\begin{gather}
\label{pdu11d}
\frac{\pd}{\pd u^1_0}\big(\breve{\mathbf{g}}(u_0,\la)\big)=
-\breve{\mathbf{g}}(u_0,\la)\cdot\cS^{-1}\Big(\frac{\pd}{\pd u^1_1}\big(\mm(u_{0},u_{1},\la)\big)
\cdot \mm(u_{0},u_{1},\la)^{-1}\Big),\\
\label{pdu21d}
\frac{\pd}{\pd u^2_0}\big(\breve{\mathbf{g}}(u_0,\la)\big)=
-\breve{\mathbf{g}}(u_0,\la)\cdot\cS^{-1}\Big(\frac{\pd}{\pd u^2_1}\big(\mm(u_{0},u_{1},\la)\big)
\cdot \mm(u_{0},u_{1},\la)^{-1}\Big)
\end{gather}
for an unknown matrix~$\breve{\mathbf{g}}(u_0,\la)$.

For the matrix~\er{Mmtoda} we have
\begin{gather}
\notag
\frac{\pd}{\pd u^1_1}\big(\mm(u_{0},u_{1},\la)\big)
\cdot \mm(u_{0},u_{1},\la)^{-1}=
\begin{pmatrix}
 \frac{1}{\cto-\ctt+u^1_{1}-u^2_{1}} & \frac{\lambda-\ctt-u^2_{1}}{\cto-\ctt+u^1_{1}-u^2_{1}} \\
 0 & 0 \\
\end{pmatrix},\\
\label{pdu11mu}
\cS^{-1}\Big(\frac{\pd}{\pd u^1_1}\big(\mm(u_{0},u_{1},\la)\big)
\cdot \mm(u_{0},u_{1},\la)^{-1}\Big)=
\begin{pmatrix}
 \frac{1}{\cto-\ctt+u^1_{0}-u^2_{0}} & \frac{\lambda-\ctt-u^2_{0}}{\cto-\ctt+u^1_{0}-u^2_{0}} \\
 0 & 0 \\
\end{pmatrix},\\
\notag
\frac{\pd}{\pd u^2_1}\big(\mm(u_{0},u_{1},\la)\big)
\cdot \mm(u_{0},u_{1},\la)^{-1}=
\begin{pmatrix}
 \frac{-1}{\cto-\ctt+u^1_{1}-u^2_{1}} & \frac{-\lambda +\cto+u^1_{1}}{\cto-\ctt+u^1_{1}-u^2_{1}} \\
 0 & 0 \\
\end{pmatrix},\\
\label{pdu21mu}
\cS^{-1}\Big(\frac{\pd}{\pd u^2_1}\big(\mm(u_{0},u_{1},\la)\big)
\cdot \mm(u_{0},u_{1},\la)^{-1}\Big)=
\begin{pmatrix}
 \frac{-1}{\cto-\ctt+u^1_{0}-u^2_{0}} & \frac{-\lambda +\cto+u^1_{0}}{\cto-\ctt+u^1_{0}-u^2_{0}} \\
 0 & 0 \\
\end{pmatrix}.
\end{gather}

In what follows the elements of the $2\times 2$ matrix~$\breve{\mathbf{g}}(u_0,\la)$ 
are denoted by~$g^{ij}(u_0,\la)$, $\,i,j=1,2$, where $u_0=(u^1_0,u^2_0)$.
Substituting~\er{pdu11mu},~\er{pdu21mu} and 
$\breve{\mathbf{g}}(u_0,\la)=\begin{pmatrix}
g^{11}(u_0,\la)&g^{12}(u_0,\la)\\
g^{21}(u_0,\la)&g^{22}(u_0,\la)
\end{pmatrix}$
to equations~\er{pdu11d},~\er{pdu21d}, we get
\begin{gather}
\label{pdu11d4}
\frac{\pd}{\pd u^1_0}\begin{pmatrix}
g^{11}(u_0,\la)&g^{12}(u_0,\la)\\
g^{21}(u_0,\la)&g^{22}(u_0,\la)
\end{pmatrix}=
-\begin{pmatrix}
g^{11}(u_0,\la)&g^{12}(u_0,\la)\\
g^{21}(u_0,\la)&g^{22}(u_0,\la)
\end{pmatrix}
\begin{pmatrix}
 \frac{1}{\cto-\ctt+u^1_{0}-u^2_{0}} & \frac{\lambda-\ctt-u^2_{0}}{\cto-\ctt+u^1_{0}-u^2_{0}} \\
 0 & 0 \\
\end{pmatrix},\\
\label{pdu21d4}
\frac{\pd}{\pd u^2_0}\begin{pmatrix}
g^{11}(u_0,\la)&g^{12}(u_0,\la)\\
g^{21}(u_0,\la)&g^{22}(u_0,\la)
\end{pmatrix}=
-\begin{pmatrix}
g^{11}(u_0,\la)&g^{12}(u_0,\la)\\
g^{21}(u_0,\la)&g^{22}(u_0,\la)
\end{pmatrix}
\begin{pmatrix}
 \frac{-1}{\cto-\ctt+u^1_{0}-u^2_{0}} & \frac{-\lambda +\cto+u^1_{0}}{\cto-\ctt+u^1_{0}-u^2_{0}} \\
 0 & 0 \\
\end{pmatrix}.
\end{gather}

It is sufficient to find one (preferably as simple as possible) invertible matrix
$$
\breve{\mathbf{g}}(u_0,\la)=\begin{pmatrix}
g^{11}(u_0,\la)&g^{12}(u_0,\la)\\
g^{21}(u_0,\la)&g^{22}(u_0,\la)
\end{pmatrix}
$$
satisfying~\er{pdu11d4} and~\er{pdu21d4}.
Equations~\er{pdu11d4},~\er{pdu21d4} allow us to take
\begin{gather}
\label{g2122}
g^{21}(u_0,\la)=0,\qquad\qquad g^{22}(u_0,\la)=1.
\end{gather}
Taking into account~\er{g2122}, we can rewrite~\er{pdu11d4},~\er{pdu21d4} as the system
\begin{gather}
\label{pdu11g11}
\frac{\pd g^{11}}{\pd u^1_0}=
\frac{-g^{11}}{\cto-\ctt+u^1_{0}-u^2_{0}},\\
\label{pdu11g12}
\frac{\pd g^{12}}{\pd u^1_0}=
\frac{-g^{11}\cdot(\lambda-\ctt-u^2_{0})}{\cto-\ctt+u^1_{0}-u^2_{0}},\\
\label{pdu21g11}
\frac{\pd g^{11}}{\pd u^2_0}=
\frac{g^{11}}{\cto-\ctt+u^1_{0}-u^2_{0}},\\
\label{pdu21g12}
\frac{\pd g^{12}}{\pd u^2_0}=
\frac{-g^{11}\cdot(-\lambda +\cto+u^1_{0})}{\cto-\ctt+u^1_{0}-u^2_{0}},
\end{gather}
where $g^{11}=g^{11}(u^1_0,u^2_0,\la)$, $\,g^{12}=g^{12}(u^1_0,u^2_0,\la)$.
It is sufficient to find some functions $g^{11}$, $\,g^{12}$ such that
$g^{11}$, $\,g^{12}$ satisfy~\mbox{\er{pdu11g11}--\er{pdu21g12}}
and the function~$g^{11}$ is nonzero 
(so that $\breve{\mathbf{g}}(u_0,\la)=\begin{pmatrix}
g^{11}&g^{12}\\
0&1
\end{pmatrix}$ is invertible).

From~\er{pdu11g11},~\er{pdu21g11} it is clear that one can take
\begin{gather}
\label{g11uu}
g^{11}(u^1_0,u^2_0,\la)=\frac{1}{\cto-\ctt+u^1_{0}-u^2_{0}}.
\end{gather}
Substituting~\er{g11uu} to equations~\er{pdu11g12},~\er{pdu21g12}, we obtain
\begin{gather}
\label{pdu10g12}
\frac{\pd g^{12}}{\pd u^1_0}=
\frac{\ctt+u^2_{0}-\lambda}{(\cto-\ctt+u^1_{0}-u^2_{0})^2},\\
\label{pdu20g12}
\frac{\pd g^{12}}{\pd u^2_0}=
\frac{\lambda-\cto-u^1_{0}}{(\cto-\ctt+u^1_{0}-u^2_{0})^2}.
\end{gather}
Integrating~\er{pdu10g12} and~\er{pdu20g12}, we see that one can take
\begin{gather}
\label{g12uu}
g^{12}(u^1_0,u^2_0,\la)=\frac{\lambda -\cto-u^1_{0}}{\cto-\ctt+u^1_{0}-u^2_{0}}.
\end{gather}

Thus we have constructed the following matrix
\begin{gather}
\label{mgu0}
\breve{\mathbf{g}}(u_0,\la)=\begin{pmatrix}
\frac{1}{\cto-\ctt+u^1_{0}-u^2_{0}} & \frac{\lambda-\cto-u^1_{0}}{\cto-\ctt+u^1_{0}-u^2_{0}}\\
0& 1
\end{pmatrix},
\end{gather}
which can be used as a gauge transformation.

Applying the gauge transformation~\er{mgu0} to the MLR~\er{Mmtoda},~\er{Umtoda}
as written in~\er{hmgu1},~\er{hugu1}, we get 
\begin{gather}
\label{tMmtoda}
\breve{\mm}(u_{0},\la)=\cS\big(\breve{\mathbf{g}}(u_0,\la)\big)\cdot \mm(u_{0},u_{1},\la)\cdot\breve{\mathbf{g}}(u_0,\la)^{-1}=
\begin{pmatrix}
 \frac{(\ctt-\cto+u^2_{0}-u^1_{0}) u^2_{0}}{u^1_{0}-u^2_{0}} & \frac{(\lambda -\cto) u^2_{0}}{u^1_{0}-u^2_{0}} \\
 \ctt-\cto+u^2_{0}-u^1_{0} & \lambda -\cto-u^1_{0} \\
\end{pmatrix},\\
\label{tUmtoda}
\begin{split}
\breve{\lt}(u_{-1},u_0,\la)&=D_t\big(\breve{\mathbf{g}}(u_0,\la)\big)\cdot\breve{\mathbf{g}}(u_0,\la)^{-1}+
\breve{\mathbf{g}}(u_0,\la)\cdot\lt(u_{-1},u_{0},u_{1},\la)\cdot\breve{\mathbf{g}}(u_0,\la)^{-1}=\\
&=\begin{pmatrix}
\lambda-\cto+\frac{u^2_{0} u^1_{-1}-u^1_{0} u^1_{-1}}{u^1_{-1}-u^2_{-1}} & \frac{(\cto-\lambda) u^2_{-1}}{u^1_{-1}-u^2_{-1}} \\
 \cto-\ctt+u^1_{0}-u^2_{0} & \frac{u^2_{-1}(\ctt-\cto+u^2_{0}-u^1_{0})}{u^1_{-1}-u^2_{-1}} \\
\end{pmatrix}.
\end{split}
\end{gather}

The fact that the matrix~\er{tMmtoda} depends only on~$u_0$~and~$\la$
allows one to construct a family of new MTs as follows.

According to Definition~\ref{dmlpgtnew}, since the matrices~\er{tMmtoda},~\er{tUmtoda} 
form an MLR for equation~\er{mtoda}, we can consider the auxiliary linear system
\begin{gather}
\lb{syspsit}
\begin{aligned}
\cS(\Psi)&=\breve{\mm}(u_{0},\la)\cdot\Psi,\\
\pd_t(\Psi)&=\breve{\lt}(u_{-1},u_0,\la)\cdot\Psi,
\end{aligned}
\end{gather}
which is compatible modulo equation~\eqref{mtoda}. 
Here $\Psi=\Psi(n,t)$ is an invertible $2\times 2$ matrix
whose elements we denote by $\uppsi^{ij}=\uppsi^{ij}(n,t)$, $\,i,j=1,2$.

Let 
$$
\breve{\mme}^{ij}=\breve{\mme}^{ij}(u_{0},\la),\qquad\quad
\breve{\lte}^{ij}=\breve{\lte}^{ij}(u_{-1},u_0,\la),\qquad\quad i,j=1,2,
$$
be the elements of the $2\times 2$ matrices 
$\hat{\mm}$, $\,\hat{\lt}$ given by~\er{tMmtoda},~\er{tUmtoda}. We have
\begin{gather}
\label{etMmtoda}
\breve{\mm}(u_{0},\la)=
\begin{pmatrix}
 \frac{(\ctt-\cto+u^2_{0}-u^1_{0}) u^2_{0}}{u^1_{0}-u^2_{0}} & \frac{(\lambda -\cto) u^2_{0}}{u^1_{0}-u^2_{0}} \\
 \ctt-\cto+u^2_{0}-u^1_{0} & \lambda -\cto-u^1_{0} 
\end{pmatrix}=
\begin{pmatrix}
\breve{\mme}^{11}(u_{0},\la) & \breve{\mme}^{12}(u_{0},\la)  \\
\breve{\mme}^{21}(u_{0},\la) & \breve{\mme}^{22}(u_{0},\la)
\end{pmatrix},\\
\label{etUmtoda}
\begin{split}
\breve{\lt}(u_{-1},u_0,\la)&=\begin{pmatrix}
\lambda-\cto+\frac{u^2_{0} u^1_{-1}-u^1_{0} u^1_{-1}}{u^1_{-1}-u^2_{-1}} & \frac{(\cto-\lambda) u^2_{-1}}{u^1_{-1}-u^2_{-1}} \\
 \cto-\ctt+u^1_{0}-u^2_{0} & \frac{u^2_{-1}(\ctt-\cto+u^2_{0}-u^1_{0})}{u^1_{-1}-u^2_{-1}} \\
\end{pmatrix}=\\
&=
\begin{pmatrix}
\breve{\lte}^{11}(u_{-1},u_0,\la) & \breve{\lte}^{12}(u_{-1},u_0,\la) \\
\breve{\lte}^{21}(u_{-1},u_0,\la) & \breve{\lte}^{22}(u_{-1},u_0,\la)
\end{pmatrix}.
\end{split}
\end{gather}
System~\er{syspsit} reads
\begin{gather}
\label{psi11t}
\begin{aligned}
\cS\begin{pmatrix}
\uppsi^{11}&\uppsi^{12}\\
\uppsi^{21}&\uppsi^{22}
\end{pmatrix}&=
\begin{pmatrix}
\breve{\mme}^{11}(u_{0},\la) & \breve{\mme}^{12}(u_{0},\la)  \\
\breve{\mme}^{21}(u_{0},\la) & \breve{\mme}^{22}(u_{0},\la)
\end{pmatrix}
\begin{pmatrix}
\uppsi^{11}&\uppsi^{12}\\
\uppsi^{21}&\uppsi^{22}
\end{pmatrix},\\
\pd_t\begin{pmatrix}
\uppsi^{11}&\uppsi^{12}\\
\uppsi^{21}&\uppsi^{22}
\end{pmatrix}&=\begin{pmatrix}
\breve{\lte}^{11}(u_{-1},u_0,\la) & \breve{\lte}^{12}(u_{-1},u_0,\la) \\
\breve{\lte}^{21}(u_{-1},u_0,\la) & \breve{\lte}^{22}(u_{-1},u_0,\la)
\end{pmatrix}\begin{pmatrix}
\uppsi^{11}&\uppsi^{12}\\
\uppsi^{21}&\uppsi^{22}
\end{pmatrix}.
\end{aligned}
\end{gather}
From~\er{psi11t} one gets
\begin{gather}
\lb{sp11}
\cS(\uppsi^{11})=\breve{\mme}^{11}\uppsi^{11}+\breve{\mme}^{12}\uppsi^{21},\qquad\quad
\cS(\uppsi^{21})=\breve{\mme}^{21}\uppsi^{11}+\breve{\mme}^{22}\uppsi^{21},\\
\lb{dp11}
\pd_t(\uppsi^{11})=\breve{\lte}^{11}\uppsi^{11}+\breve{\lte}^{12}\uppsi^{21},\qquad\quad
\pd_t(\uppsi^{21})=\breve{\lte}^{21}\uppsi^{11}+\breve{\lte}^{22}\uppsi^{21}.
\end{gather}
We set $\vv=\dfrac{\uppsi^{11}}{\uppsi^{21}}$.
From~\er{sp11},~\er{dp11} we obtain
\begin{gather}
\label{swp11}
\cS(\vv)=\frac{\cS(\uppsi^{11})}{\cS(\uppsi^{21})}=
\frac{\breve{\mme}^{11}\uppsi^{11}+\breve{\mme}^{12}\uppsi^{21}}{\breve{\mme}^{21}\uppsi^{11}+\breve{\mme}^{22}\uppsi^{21}}=
\frac{\dfrac{\breve{\mme}^{11}\uppsi^{11}}{\uppsi^{21}}+\breve{\mme}^{12}}{\dfrac{\breve{\mme}^{21}\uppsi^{11}}{\uppsi^{21}}+\breve{\mme}^{22}}=
\frac{\breve{\mme}^{11}\vv+\breve{\mme}^{12}}{\breve{\mme}^{21}\vv+\breve{\mme}^{22}},\\
\label{dwp11}
\begin{split}
\pd_t(\vv)&=\pd_t\Big(\frac{\uppsi^{11}}{\uppsi^{21}}\Big)=
\frac{\pd_t(\uppsi^{11})}{\uppsi^{21}}-\frac{\uppsi^{11}\pd_t(\uppsi^{21})}{(\uppsi^{21})^2}=
\frac{\breve{\lte}^{11}\uppsi^{11}+\breve{\lte}^{12}\uppsi^{21}}{\uppsi^{21}}-
\frac{\uppsi^{11}\big(\breve{\lte}^{21}\uppsi^{11}+\breve{\lte}^{22}\uppsi^{21}\big)}{(\uppsi^{21})^2}=\\
&=\breve{\lte}^{11}\frac{\uppsi^{11}}{\uppsi^{21}}+\breve{\lte}^{12}
-\breve{\lte}^{21}\Big(\frac{\uppsi^{11}}{\uppsi^{21}}\Big)^2-
\breve{\lte}^{22}\frac{\uppsi^{11}}{\uppsi^{21}}=
-\breve{\lte}^{21}\vv^2+(\breve{\lte}^{11}-\breve{\lte}^{22})\vv+\breve{\lte}^{12}.
\end{split}
\end{gather}
Since, according to~\er{etMmtoda},~\er{etUmtoda}, for $i,j=1,2$ one has
$$
\breve{\mme}^{ij}=\breve{\mme}^{ij}(u^1_{0},u^2_{0},\la),\qquad\quad
\breve{\lte}^{ij}=\breve{\lte}^{ij}(u^1_{-1},u^2_{-1},u^1_{0},u^2_{0},\la),
$$
equations~\er{swp11},~\er{dwp11} can be written as
\begin{gather}
\label{swuu}
\cS(\vv)=\frac{\breve{\mme}^{11}(u^1_{0},u^2_{0},\la)\vv+\breve{\mme}^{12}(u^1_{0},u^2_{0},\la)}%
{\breve{\mme}^{21}(u^1_{0},u^2_{0},\la)\vv+\breve{\mme}^{22}(u^1_{0},u^2_{0},\la)},\\
\label{dwuu}
\begin{split}
\pd_t&(\vv)=-\breve{\lte}^{21}(u^1_{-1},u^2_{-1},u^1_{0},u^2_{0},\la)\vv^2+\\
&+\big(\breve{\lte}^{11}(u^1_{-1},u^2_{-1},u^1_{0},u^2_{0},\la)
-\breve{\lte}^{22}(u^1_{-1},u^2_{-1},u^1_{0},u^2_{0},\la)\big)\vv
+\breve{\lte}^{12}(u^1_{-1},u^2_{-1},u^1_{0},u^2_{0},\la),
\end{split}
\end{gather}
where the parameter~$\la$ can take any value from~$\fik$.

Fix constants $\la_1,\la_2\in\fik$, $\,\la_1\neq\la_2$.
Considering equations~\er{swuu},~\er{dwuu} in the case~$\la=\la_1$ 
and denoting~$\vv$ by~$w^1$ in this case, we obtain
\begin{gather}
\label{1swuu}
\cS(w^1)=\frac{\breve{\mme}^{11}(u^1_{0},u^2_{0},\la_1)w^1+\breve{\mme}^{12}(u^1_{0},u^2_{0},\la_1)}%
{\breve{\mme}^{21}(u^1_{0},u^2_{0},\la_1)w^1+\breve{\mme}^{22}(u^1_{0},u^2_{0},\la_1)},\\
\label{1dwuu}
\begin{split}
\pd_t&(w^1)=-\breve{\lte}^{21}(u^1_{-1},u^2_{-1},u^1_{0},u^2_{0},\la_1)(w^1)^2+\\
&+\big(\breve{\lte}^{11}(u^1_{-1},u^2_{-1},u^1_{0},u^2_{0},\la_1)
-\breve{\lte}^{22}(u^1_{-1},u^2_{-1},u^1_{0},u^2_{0},\la_1)\big)w^1
+\breve{\lte}^{12}(u^1_{-1},u^2_{-1},u^1_{0},u^2_{0},\la_1).
\end{split}
\end{gather}
Similarly, considering equations~\er{swuu},~\er{dwuu} in the case~$\la=\la_2$ 
and denoting~$\vv$ by~$w^2$ in this case, we get
\begin{gather}
\label{2swuu}
\cS(w^2)=\frac{\breve{\mme}^{11}(u^1_{0},u^2_{0},\la_2)w^2+\breve{\mme}^{12}(u^1_{0},u^2_{0},\la_2)}%
{\breve{\mme}^{21}(u^1_{0},u^2_{0},\la_2)w^2+\breve{\mme}^{22}(u^1_{0},u^2_{0},\la_2)},\\
\label{2dwuu}
\begin{split}
\pd_t&(w^2)=-\breve{\lte}^{21}(u^1_{-1},u^2_{-1},u^1_{0},u^2_{0},\la_2)(w^2)^2+\\
&+\big(\breve{\lte}^{11}(u^1_{-1},u^2_{-1},u^1_{0},u^2_{0},\la_2)
-\breve{\lte}^{22}(u^1_{-1},u^2_{-1},u^1_{0},u^2_{0},\la_2)\big)w^2
+\breve{\lte}^{12}(u^1_{-1},u^2_{-1},u^1_{0},u^2_{0},\la_2).
\end{split}
\end{gather}

As usual, we are going to use the notation
\begin{gather}
\label{w12k}
w^1_\ik=\cS^\ik(w^1),\qquad\quad
w^2_\ik=\cS^\ik(w^2)\qquad\quad\forall\,\ik\in\zz.
\end{gather}

To avoid cumbersome formulas, we fix a constant $\hc\in\fik$ and consider the case
\begin{gather}
\label{ct0}
\cto=\ctt=0,\qquad\quad\la_1=\hc,\qquad\quad\la_2=0.
\end{gather}
In the case~\er{ct0}, from equations \er{1swuu} and~\er{2swuu}
we can express $u^1_{0}$, $u^2_{0}$ in terms of
$$
w^1_0=w^1,\qquad w^2_0=w^2,\qquad w^1_1=\cS(w^1),\qquad w^2_1=\cS(w^2)
$$
as follows 
\begin{gather}
\lb{mthw}
\left\{
\begin{gathered}
u^1=-\Big(\hc w^{2}_{0} (1+w^{2}_{1}) 
(-w^{1}_{1} w^{2}_{0}+w^{2}_{1}+w^{1}_{1} w^{2}_{1}+\\
+w^{2}_{0} w^{2}_{1})\Big)\left/\Big((w^{2}_{0}-w^{2}_{1}) 
(w^{1}_{1} w^{2}_{0}+w^{1}_{0} w^{1}_{1} w^{2}_{0}\right.\\
-w^{1}_{0} w^{2}_{1}-w^{1}_{0} w^{1}_{1} w^{2}_{1}-w^{1}_{0} w^{2}_{0} w^{2}_{1}+w^{1}_{1} w^{2}_{0} w^{2}_{1})\Big),\\
u^2=-\Big(\hc (1+w^{2}_{0}) w^{2}_{1} (-w^{1}_{1} w^{2}_{0}+w^{2}_{1}+w^{1}_{1} w^{2}_{1}+\\
+w^{2}_{0} w^{2}_{1})\Big)\left/\Big((w^{2}_{0}-w^{2}_{1}) (w^{1}_{1} w^{2}_{0}+w^{1}_{0} w^{1}_{1} w^{2}_{0}\right.\\
-w^{1}_{0} w^{2}_{1}-w^{1}_{0} w^{1}_{1} w^{2}_{1}-w^{1}_{0} w^{2}_{0} w^{2}_{1}+w^{1}_{1} w^{2}_{0} w^{2}_{1})\Big).
\end{gathered}\right.
\end{gather}
For each $\ik\in\zz$, applying the operator~$\cS^\ik$ to~\er{mthw}, we get
\begin{gather}
\lb{uksfw}
\left\{
\begin{gathered}
u^1_\ik=-\Big(\hc w^{2}_{\ik} (1+w^{2}_{\ik+1}) 
(-w^{1}_{\ik+1} w^{2}_{\ik}+w^{2}_{\ik+1}+w^{1}_{\ik+1} w^{2}_{\ik+1}+\\
+w^{2}_{\ik} w^{2}_{\ik+1})\Big)\left/\Big((w^{2}_{\ik}-w^{2}_{\ik+1}) 
(w^{1}_{\ik+1} w^{2}_{\ik}+w^{1}_{\ik} w^{1}_{\ik+1} w^{2}_{\ik}\right.\\
-w^{1}_{\ik} w^{2}_{\ik+1}-w^{1}_{\ik} w^{1}_{\ik+1} w^{2}_{\ik+1}-w^{1}_{\ik} w^{2}_{\ik} w^{2}_{\ik+1}+w^{1}_{\ik+1} w^{2}_{\ik} w^{2}_{\ik+1})\Big),\\
u^2_\ik=-\Big(\hc (1+w^{2}_{\ik}) w^{2}_{\ik+1} (-w^{1}_{\ik+1} w^{2}_{\ik}+w^{2}_{\ik+1}+w^{1}_{\ik+1} w^{2}_{\ik+1}+\\
+w^{2}_{\ik} w^{2}_{\ik+1})\Big)\left/\Big((w^{2}_{\ik}-w^{2}_{\ik+1}) (w^{1}_{\ik+1} w^{2}_{\ik}+w^{1}_{\ik} w^{1}_{\ik+1} w^{2}_{\ik}\right.\\
-w^{1}_{\ik} w^{2}_{\ik+1}-w^{1}_{\ik} w^{1}_{\ik+1} w^{2}_{\ik+1}-w^{1}_{\ik} w^{2}_{\ik} w^{2}_{\ik+1}+w^{1}_{\ik+1} w^{2}_{\ik} w^{2}_{\ik+1})\Big),\\
\ik\in\zz.
\end{gathered}\right.
\end{gather}

Substituting~\er{uksfw} in the right-hand sides of formulas~\er{1dwuu},~\er{2dwuu}, 
we obtain the two-component equation
\begin{gather}
\lb{eqhw}
\left\{
\begin{gathered}
\pd_t(w^1)=-\Big(\hc (w^{1}_{1} (w^{2}_{0})^2+w^{1}_{0} w^{1}_{1} (w^{2}_{0})^2+w^{1}_{1} w^{2}_{-1} (w^{2}_{0})^2+\\
+w^{1}_{0} w^{1}_{1} w^{2}_{-1} (w^{2}_{0})^2-w^{1}_{0} w^{2}_{-1} w^{2}_{1}-w^{1}_{0} w^{1}_{1} w^{2}_{-1} w^{2}_{1}\\
-2 w^{1}_{0} w^{2}_{-1} w^{2}_{0} w^{2}_{1}-2 w^{1}_{0} w^{1}_{1} w^{2}_{-1} w^{2}_{0} w^{2}_{1}+\\
+w^{1}_{1} (w^{2}_{0})^2 w^{2}_{1}+w^{1}_{0} w^{1}_{1} (w^{2}_{0})^2 w^{2}_{1}-w^{1}_{0} w^{2}_{-1} (w^{2}_{0})^2 w^{2}_{1}+\\
+w^{1}_{1} w^{2}_{-1} (w^{2}_{0})^2 w^{2}_{1})\Big)\left/\Big((w^{2}_{-1}-w^{2}_{0}) (w^{1}_{1} w^{2}_{0}+\right.\\
+w^{1}_{0} w^{1}_{1} w^{2}_{0}-w^{1}_{0} w^{2}_{1}-w^{1}_{0} w^{1}_{1} w^{2}_{1}-w^{1}_{0} w^{2}_{0} w^{2}_{1}+w^{1}_{1} w^{2}_{0} w^{2}_{1})\Big),\\
\pd_t(w^2)=-\Big(\hc w^{2}_{0} (1+w^{2}_{0}) (w^{1}_{1} w^{2}_{0}-w^{2}_{1}-w^{1}_{1} w^{2}_{1}\\
-w^{2}_{0} w^{2}_{1})\Big)\left/\Big(w^{1}_{1} w^{2}_{0}
+w^{1}_{0} w^{1}_{1} w^{2}_{0}-w^{1}_{0} w^{2}_{1}\right.\\
-w^{1}_{0} w^{1}_{1} w^{2}_{1}-w^{1}_{0} w^{2}_{0} w^{2}_{1}+w^{1}_{1} w^{2}_{0} w^{2}_{1}\Big).
\end{gathered}\right.
\end{gather}

Recall that system~\er{syspsit} is compatible modulo equation~\eqref{mtoda}
and is equivalent to~\er{sp11},~\er{dp11}. 
Since equations \er{swuu}, \er{dwuu}, \er{1swuu}, \er{1dwuu}, \er{2swuu}, \er{2dwuu}
are derived from~\er{sp11},~\er{dp11}, 
system~\er{1swuu},~\er{1dwuu} and system~\er{2swuu},~\er{2dwuu}
are compatible modulo~\eqref{mtoda} as well.
Since \er{mthw},~\er{eqhw} are derived from
\er{1swuu}, \er{1dwuu}, \er{2swuu}, \er{2dwuu} with~\er{ct0}, 
system~\er{mthw},~\er{eqhw} is also compatible modulo equation~\eqref{mtoda} with $\cto=\ctt=0$.
This implies that \er{mthw} is an MT from~\er{eqhw} to~\eqref{mtoda}
with $\cto=\ctt=0$.

Computing the composition of the MTs~\er{mthw} and~\er{dstoda},
one gets the following MT 
\begin{gather}
\lb{tdw}
\left\{
\begin{gathered}
\td^1=-\Big(\hc^2 w^{2}_{0} (1+w^{2}_{0}) w^{2}_{1} (1+w^{2}_{1}) (-w^{1}_{1} w^{2}_{0}+w^{2}_{1}+\\
+w^{1}_{1} w^{2}_{1}+w^{2}_{0} w^{2}_{1}) (-w^{1}_{2} w^{2}_{1}+w^{2}_{2}+w^{1}_{2} w^{2}_{2}+\\
+w^{2}_{1} w^{2}_{2})\Big)\left/\Big((w^{2}_{0}-w^{2}_{1})^2 (w^{1}_{1} w^{2}_{0}+\right.\\
+w^{1}_{0} w^{1}_{1} w^{2}_{0}-w^{1}_{0} w^{2}_{1}-w^{1}_{0} w^{1}_{1} w^{2}_{1}-w^{1}_{0} w^{2}_{0} w^{2}_{1}+\\
+w^{1}_{1} w^{2}_{0} w^{2}_{1}) (-w^{1}_{2} w^{2}_{1}-w^{1}_{1} w^{1}_{2} w^{2}_{1}+w^{1}_{1} w^{2}_{2}+\\
+w^{1}_{1} w^{1}_{2} w^{2}_{2}+w^{1}_{1} w^{2}_{1} w^{2}_{2}-w^{1}_{2} w^{2}_{1} w^{2}_{2})\Big),\\
\td^2=-\Big(\hc w^{2}_{1} (1+w^{2}_{1}) (w^{2}_{0}-w^{2}_{2}) (-w^{1}_{2} w^{2}_{1}+w^{2}_{2}+w^{1}_{2} w^{2}_{2}+\\
+w^{2}_{1} w^{2}_{2})\Big)\left/\Big((w^{2}_{0}-w^{2}_{1}) (w^{2}_{1}-w^{2}_{2}) (-w^{1}_{2} w^{2}_{1}-w^{1}_{1} w^{1}_{2} w^{2}_{1}+\right.\\
+w^{1}_{1} w^{2}_{2}+w^{1}_{1} w^{1}_{2} w^{2}_{2}+w^{1}_{1} w^{2}_{1} w^{2}_{2}-w^{1}_{2} w^{2}_{1} w^{2}_{2})\Big).
\end{gathered}\right.
\end{gather}
from equation~\er{eqhw} to equation~\er{todanew}.

\begin{remark}
\lb{rinteqhw}
As shown above, the matrices~\er{tMmtoda},~\er{tUmtoda} 
form a $\la$-dependent MLR for~\er{mtoda}, 
while formula~\er{mthw} is an MT from~\er{eqhw} 
to equation~\eqref{mtoda} with $\cto=\ctt=0$.
Similarly to Remark~\ref{rintdtv}, these two facts yield the following.
Substituting~\er{mthw} in~\er{tMmtoda},~\er{tUmtoda} with $\cto=\ctt=0$, 
one obtains a $\la$-dependent MLR for equation~\er{eqhw}.  
This MLR for~\er{eqhw} seems to be new.

Equation~\er{eqhw} is integrable in the sense that 
\er{eqhw} has a nontrivial MLR with parameter~$\la$
and is connected by the MT~\er{tdw} 
to the well-known integrable Toda lattice 
(in the Flaschka--Manakov coordinates) equation~\er{todanew}.
\end{remark}
The obtained equation~\er{eqhw} and MTs~\er{mthw},~\er{tdw} 
are new, to our knowledge.

\section*{Acknowledgments}

The work on the results presented in Section~\ref{sbc}
was funded by the Russian Science Foundation project No. 20-71-10110 (\url{https://rscf.ru/en/project/23-71-50012/}).

The work on Section~\ref{stoda}
was done with financial support by the Ministry of Science and Higher Education of the Russian Federation
(Agreement No.~{075-02-2024-1442})
within the activity of the Regional Mathematical Center of the P.G. Demidov Yaroslavl State University.

\end{document}